%% file: main.tex
\documentclass[twocolumn,numberedappendix,apj]{openjournal}
\input{Packages/packages.tex}

\begin{document}

\newcommand{\cmt}[1]{\textcolor{red}{\textsf{#1}}}
\newcommand{\REV}[1]{\textcolor{red}{$\langle$\textsf{Rev: #1}$\rangle$}}
\newcommand{\CE}[1]{\textcolor{cyan}{$\langle$\textsf{CE: #1}$\rangle$}}
\newcommand{\GC}[1]{\textcolor{green}{$\langle$\textsf{GC: #1}$\rangle$}}

\defcitealias{PaperII}{Paper II}
\defcitealias{PaperIII}{Paper III}

\title[]{DIPLODOCUS I: Framework for the evaluation of relativistic transport equations with continuous forcing and discrete particle interactions}
\shorttitle{DIPLODOCUS I}
\author{\vspace{-1.5cm}Christopher N. Everett$^{1,\star}$\orcidlink{0000-0001-5181-108X}, Garret Cotter$^{1}$\orcidlink{0000-0002-9975-1829}}
\affiliation{$^1$Oxford Astrophysics, Denys Wilkinson Building, Keble Road, Oxford, OX1 3RH, United Kingdom}
\thanks{$^{\star}$Contact e-mail:  \href{mailto:christopher.everett@physics.ox.ac.uk}{christopher.everett@physics.ox.ac.uk}}



\begin{abstract}
    \input{Sections/abstract.tex}
\end{abstract}


\maketitle

\input{Sections/introduction.tex}
\input{Sections/BE_general.tex}

\input{Sections/BE_coordinate.tex}

\input{Sections/DIP.tex}

\input{Sections/conclusion.tex}

\section*{Acknowledgements}
\vspace{3pt}

C.\,N.\,Everett acknowledges a Science and Technologies Facilities Council (STFC) studentship ST/X508664/1. G.\,Cotter acknowledges support from STFC grants ST/V006355/1, ST/V001477/1, and ST/S002952/1, and from Exeter College, Oxford. The authors would also like to thank James Matthews, Marc Klinger-Plaisier, Sera Markoff (and their research group), Pedro Ferreira, and Dion Everett for their helpful insights during various phases of this work.

\section*{Data Availability}
\vspace{3pt}
No data were created or analysed in this article.

\bibliographystyle{mnras_modified}
\bibliography{references}

\appendix
\input{Appendicies/absorptionanddecay.tex}

\input{Appendicies/metrics.tex}
\input{Appendicies/longequations.tex}

\end{document}

%% file: Packages/packages.tex
\usepackage[utf8]{inputenc}
\usepackage[T1]{fontenc}
\usepackage{ae,aecompl}

\usepackage{graphicx}
\graphicspath{Figures/}
\usepackage{amsmath,amsthm,amssymb,amsfonts,amstext}
\usepackage{bm}
\usepackage{orcidlink}

\usepackage{enumitem}

\newcommand{\aref}[1]{Appendix~\ref{#1}}

\newcommand{\eref}[1]{Eq.~(\ref{#1})}
\newcommand{\esref}[1]{Eqs.~(\ref{#1})}
\newcommand{\fref}[1]{Fig.~\ref{#1}}

\newcommand{\sref}[1]{Section~\ref{#1}}
\newcommand{\ssref}[1]{Sections~\ref{#1}}

\usepackage{hyperref}
\hypersetup{colorlinks=true,linkcolor=blue,citecolor=blue,filecolor=blue,urlcolor=blue}
\usepackage[all]{hypcap}

%% file: Sections/abstract.tex
DIPLODOCUS (\textbf{D}istribution-\textbf{I}n-\textbf{PL}ateaux meth\textbf{ODO}logy for the \textbf{C}omp\textbf{U}tation of transport equation\textbf{S}) is a framework being developed for the mesoscopic modelling of astrophysical systems via the transport of particle distribution functions through the seven dimensions of phase space, including continuous forces and discrete interactions between particles. This first paper in a series provides an overview of the analytical framework behind the model, consisting of an integral formulation of the relativistic transport equations (Boltzmann equations) and a discretisation procedure for the particle distribution function (\textit{Distribution-In-Plateaux}). The latter allows for the evaluation of anisotropic interactions and generates a conservative numerical scheme for a distribution function's transport through phase space.

%% file: Sections/introduction.tex
\section{Introduction}\label{sec:intro}

Astronomical phenomena must usually be inferred by distant observation alone. Blazars are a class of objects whose observed emissions are thought to originate from magnetised, relativistic jets of material formed close to the black holes at the centre of Active Galactic Nuclei (AGN) and directed almost exactly along our line of sight. The spectra of such objects are truly multi-wavelength, spanning from radio to ultra-high-energy gamma rays, typically with a double-peaked shape \citep{UrryPadovani_1995,PadovaniEtAl_2017,BlandfordEtAl_2019,Talvikki_2019}. 

The processes that dominate this emission are generally thought to be synchrotron radiation and inverse Compton scattering from either leptonic or hadronic populations, or a combination of both, within the jets. The observed brightness of such processes is heavily dependent on the properties of the in situ particle populations; it is therefore of critical importance that such processes and populations are well modelled. 

In deriving macroscopic (hydrodynamic) models such as (General Relativistic Magneto-)HydroDynamics (GRM)HD, particle populations are assumed to be highly collisional and in local thermal equilibrium. Therefore, knowledge of their distribution in momentum space may be neglected in favour of a small set of defining parameters, e.g. temperature, density and bulk flow velocity. This reduction of dimensionality permits large spatial domains to be modelled, leading to high resolution simulations of the global jetted AGN structure \citep{McKinneyGammie_2004,McKinney_2006,ChatterjeeEtAl_2019a,WhiteheadMatthews_2023} at the cost of limited information on the underlying micro-scale interactions between particle populations, hence poor direct simulation of emissions.

By contrast, microscopic (kinetic) models maintain information about the underlying distribution of particles in momentum space, at the cost of the increased dimensionality. To make this computationally tractable, other assumptions, e.g. isotropy in momentum, are often made to reduce the dimensionality as much as possible. 

One frequently used microscopic approach, in the field of plasma physics, is that of Particle-In-Cell (PIC). PIC replaces particle distributions with a finite number of discrete particles that attempt to simulate the entire distribution. This approach works well when distributions are near thermal or when small scale plasma instabilities and dynamics require resolution, such as for particle acceleration in magnetised shock fronts and reconnection regions \citep{CeruttiEtAl_2013,SironiEtAl_2015,MehlhaffEtAl_2024,ComissoEtAl_2024}. However, due to computational limits, PIC can only simulate a finite number of particles. This limits sampling of both the distribution and interaction processes and constrains the spatial extent of any simulation. Therefore, PIC cannot typically be applied on a macroscopic scale.

An alternative microscopic approach is to assume symmetries in the particle distribution in space and momentum space. For example, an emitting region of an astrophysical source can be simplified to a spherical or slab-like zone with an internally isotropic distribution of particles. Early astrophysical use of these methods was in modelling cosmic ray transport \citep{Lindquist_1966,Webb_1985,Webb_1989a}. Here, in addition to spherical symmetry in momentum space, specifically with reference to a co-moving frame, particle distributions were taken to be close to Maxwellian. The assumption of isotropy in particular led to extensive work \citep[for example][]{BlumenthalGould_1970,Weaver_1976,AharonianAtoyan_1981,Svensson_1982,Baring_1987,Dermer_1984,Brinkmann_1984,Dermer_1985,Dermer_1986,CoppiBlandford_1990,SarkarEtAl_2019} describing a wide range of interactions between particles and fields, with only small ventures into the world of anisotropy for specific  \citep{StepneyGuilbert_1983,MoskalenkoStrong_2000,KelnerEtAl_2014,LaiNg_2023} and general \citep{EverettCotter_2024} interactions. These works form the foundation of modern single/multi-zone models for jetted astrophysical sources \citep{MastichiadisKirk_1995,MastichiadisEtAl_2005,DimitrakoudisEtAl_2012a,PotterCotter_2012,PotterCotter_2013a,PotterCotter_2013b,PotterCotter_2013c,BottcherEtAl_2013,CerrutiEtAl_2015,GaoEtAl_2017,Potter_2018,JimenezFernandezvanEerten_2021,LucchiniEtAl_2022,Boula_2022,ZachariasEtAl_2022,HahnEtAl_2022,KlingerEtAl_2024a,StathopoulosEtAl_2024,CerrutiEtAl_2024}, which maintain the tradition of momentum-space isotropy to limit computational requirements. These requirements are further reduced by considering additional symmetries in the spatial distribution of particles. In general, such models operate by considering independent, homogeneous zones of isotropic plasma, normally spherical or slab-like in shape, with prescribed internal properties such as bulk Lorentz factor, magnetic field strength and initial particle distributions. These zones may then be placed along the axis of the jet to generate its structure, though they typically do not interact with one another. The distributions of particles within each zone are then evolved according to their interactions and the parameters of the zone, with emission spectra generated by Doppler boosting escaping photons from each zone into an observer frame. These spectra can then be fit to observational data of astrophysical sources. These zoned model fits have provided great insight into the environment and dynamics of sources like AGN jets, allowing the study of questions such as how particles are accelerated to the initial distributions, and whether the jet material is dominated by leptonic or hadronic particles, questions remain a matter of debate \citep{Weekes_2003,HoerbeEtAl_2020,KantzasEtAl_2023}. 

The insight these zoned models have provided into blazars, and jetted astrophysical sources in general, such as X-Ray Binaries (XRB) and Gamma-Ray Bursts (GRBs), should not be understated \citep[see, for example][]{MarkoffEtAl_2001,BottcherEtAl_2013,Marscher_2014,PotterCotter_2015,Potter_2018,KantzasEtAl_2021a,KlingerEtAl_2024}. However, zoned models are intrinsically limited by their constraining symmetries and it is yet to be seen if their results are robust upon relaxation of these assumptions.

Recent studies in the fields of pair-plasma winds \citep{AksenovEtAl_2004,AksenovEtAl_2007,AksenovEtAl_2010} and cosmic-ray propagation \citep{KrumholzEtAl_2022,KrumholzEtAl_2024} have demonstrated self-consistent evolution of axisymmetric particle distributions using microscopic approaches. Therefore, the additional dimensionality that comes with relaxing isotropy is within the capability of modern computing. Further, the study of the globular cluster Terzan 5 by \citet{KrumholzEtAl_2024} provided a unique explanation for inconsistent gamma-ray emissions. Low-energy gamma rays are observed coincident with the location of the globular cluster, while a source of high-energy gamma rays is spatially displaced in the direction of Terzan 5's magnetotail. This displacement was successfully modelled by the slow isotropisation of an accelerated population of cosmic rays propagating down the magnetotail. Initially, due to their narrow range of pitch angles, their emissions are beamed away from the line of sight and only once the distribution becomes sufficient isotropic do the emissions become visible. This explanation of the observed spatial separation was only made possible by considering the anisotropic evolution of the cosmic-ray population.

These findings raise many questions regarding the assumed momentum-space isotropy and spatial homogeneity of zoned models. If these assumptions were to be eliminated, would the simulated emissions continue to fit the observed data? Would this require a reinterpretation of the source, and could this provide deeper physical insights? These questions motivated the present work, which derives and details a framework designed to address them: DIPLODOCUS.

With the elimination of assumed symmetries, kinetic models boil down to a question of particle transport: how does a particle go from point $A$ to point $B$? (With the slight complication that point $A$ and point $B$ exist not in three dimensions of space but the seven dimensions of phase space that include space, time and momentum.) 

This general approach to particle transport includes aspects of both microscopic and macroscopic models; small-scale interactions and large-scale structure. As such it lies between the two---in the mesoscopic regime. This regime, and the generality of the framework, permit its application to whole host of astrophysical phenomena, not just jetted sources.

This first paper (Paper I) in the series aims to give an overarching description of this framework for mesoscopic particle transport in a very general sense, forming the backbone of this work. \sref{sec:general} describes, in a coordinate-free way, how particles get from $A$ to $B$ in phase space, and what exactly that means. \sref{sec:coordinate} then adds in coordinates to the description, and finally, \sref{sec:DIP} describes the methodology used to evaluate the derived transport equations. \citetalias{PaperII} will then describe the numerical implementation of the framework defined in Paper I, with a demonstration of  microscopic capabilities, i.e. particle, interactions, emission and forcing. \citetalias{PaperIII} will then examine the frameworks macroscopic capabilities, with the inclusion of spatial structure and advection.   

%% file: Sections/BE_general.tex
\section{Coordinate-Free Transport Equation}\label{sec:general}

This section defines (with some rigour) a coordinate-free, covariant form of the transport equations for particle distribution functions through phase space, also more generally known as Boltzmann equations \citep{Boltzmann_1872}. The use of a coordinate-free approach is mathematically compact, and allows the transport equations to be adapted to any physical (or even non-physical) systems.

In the context of astrophysical sources, particles encounter continuous conservative (Hamiltonian) forces, such as the electromagnetic Lorentz force, as well as non-conservative (non-Hamiltonian) forces, such as radiation reaction that generates synchrotron emission. Particles may also encounter discrete, discontinuous interactions, i.e. collisions, which affect their path through phase space. 

There is a rich history of developing Boltzmann equations for the purpose of astrophysical modelling. However, most derivations focus solely on conservative cases, neglecting non-conservative elements. This work primarily builds on the approaches taken by \citet{Lindquist_1966,EhlersJ_1971,GrootEtAl_1980}. Other works, some derived from the above sources and some independent, that have made impressions on the work presented here include \citet{Bichteler_1965,Hakim_1967,Hakim_1967a,Stewart_1971,Steeb_1979,Steeb_1980,EllisEtAl_1983,EllisEtAl_1983a,BlandfordEichler_1987,Cercignani_2002,CardallMezzacappa_2003,Lifshits_2008,ShibataEtAl_2014}. 

\subsection{A to B: Phase Flow on a Manifold}\label{subsec: phase flow w o coords}
\begin{figure}[!ht]
    \centering
    \includegraphics[scale=1.00]{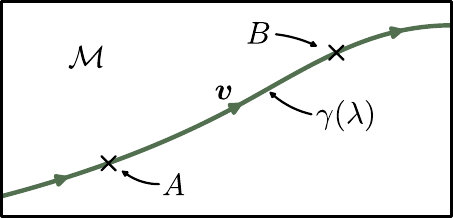}
    \vspace{5pt}
    \caption{\label{fig: A B worldline}The worldline of a single particle through phase space ($\mathcal{M}$) between two points $A$ and $B$, defined by a curve $\gamma(\lambda)$, with $\lambda$ being some affine parameter of that curve, and tangent to the vector $\bm{v}$.}
\end{figure}

As a first step, consider a single particle located at some point $A$ at some instance and at $B$ at a later instance. These points $A$ and $B$ exist in an abstract phase space, a differentiable manifold of dimension $n$, denoted by $\mathcal{M}$. Both points exist along a unique curve $\gamma(\lambda)$, where $\lambda$ is an affine parameter along that curve such that $\gamma(\lambda_A)=A$ and $\gamma(\lambda_B)=B$. The coordinates of any point along this curve have the form $\bm{y}=\left(y^1,...,y^n\right)$ and there exists a unique vector $\bm{v}$ tangent to the curve at each point. The curve $\gamma(\lambda)$ is therefore defined by the equation of motion   
\begin{equation}
    \frac{\mathrm{d}\bm{y}}{\mathrm{d}\lambda}=\bm{v},
\end{equation}
and is known as the particle's worldline through phase space (see \fref{fig: A B worldline}).  

\begin{figure}[!ht]
    \centering
    \includegraphics[scale=1.00]{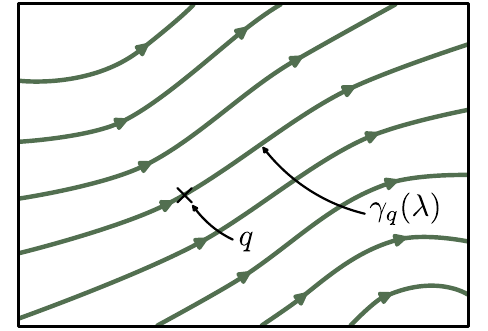}
    \vspace{5pt}
    \caption{\label{fig: A B congruence}A congruence of worldlines $\gamma_{q}$ through phase space $\mathcal{M}$ defined by the vector field $\bm{v}$ at every point $q\in\mathcal{M}$.}
\end{figure}

Now consider expanding this concept to every point $q\in\mathcal{M}$, i.e. at every point there exists a vector field $\bm{v}$ which defines a unique congruence of worldlines $\gamma_q(\lambda)$ that fill $\mathcal{M}$. These curves define how any particle will evolve if placed at any point $q\in\mathcal{M}$ (see \fref{fig: A B congruence}).

The manifold $\mathcal{M}$ has a volume form (volume element) associated with it defined by 
\begin{equation}
    \bm{\Omega}=\Omega(\bm{y})\bm{\mathrm{d}}y^1\wedge...\wedge\bm{\mathrm{d}}y^n=\Omega(\bm{y})\bm{\mathrm{d}}y^{1...n},
\end{equation}
where $\wedge$ is the wedge product and $\bm{\mathrm{d}}y^{1...n}$ is shorthand for $\bm{\mathrm{d}}y^1\wedge...\wedge\bm{\mathrm{d}}y^n$. The evolution of this volume form with the phase flow generated by $\bm{v}$ is given by a Lie derivative with respect to $\bm{v}$ (also known as the Liouville operator or vector) 
\begin{equation}
    \mathcal{L}_{\bm{v}}\bm{\Omega}=\bm{\mathrm{d}}(\bm{v}\lrcorner\bm{\Omega})+\bm{v}\lrcorner(\bm{\mathrm{d}}\bm{\Omega}),
\end{equation}
where Cartan's magic formula\footnote{The ``magic formula'' is often associated with \'Elie Cartan, who demonstrates a proof in \citet{Cartan_1922}; however \citet{Helein_2023} points out that formula also appears in the earlier work by Th\'eophile De Donder \citet{Donder_1911} (verified by the authors), making this a potential case for Stigler's law of eponymy \citep{Stigler_1980}. Hence it is unclear who it should properly be associated with.} has been used, with $\bm{\mathrm{d}}$ being the exterior derivative and $\lrcorner$ being the interior product\footnote{The interior product is not to be confused with the inner product.}. In general, as $\bm{\Omega}$ is a volume form, i.e. is a differential form of dimension $n$ on a $n$ dimensional manifold, $\bm{\mathrm{d}}\bm{\Omega}=0$ and what remains is 
\begin{equation}
    \mathcal{L}_{\bm{v}}\bm{\Omega}=\bm{\mathrm{d}}(\bm{v}\lrcorner\bm{\Omega})=(\text{div}_{\bm{\Omega}}\bm{v})\bm{\Omega}=\bm{\mathrm{d}}\bm{\omega},
\end{equation}
where $\text{div}_{\bm{\Omega}}$ is the $\bm{\Omega}$-divergence. Note that $\bm{v}\lrcorner\bm{\Omega}\equiv\bm{\omega}$ defines the hypersurface element for hypersurfaces in $\mathcal{M}$.

\subsection{The Distribution Function}\label{subsec: distribution function}

For Hamiltonian (conservative) systems, $\bm{\mathrm{d}}\bm{\omega}=0$ and therefore $\mathcal{L}_{\bm{v}}\bm{\Omega}=0$, which is a statement of Liouville's theorem \citep{Liouville_1838} that the \textit{phase space volume is invariant with respect to the phase flow}. Systems of importance to astrophysics are not in general Hamiltonian (see \sref{subsec: phase flow w coords}) and as such the phase space volume is \textit{not}, in general, invariant with respect to the phase flow i.e. $\mathcal{L}_{\bm{v}}\bm{\Omega}\neq0$. Hence Liouville's theorem does not apply and instead must be generalised.

Consider a new volume form $\bar{\bm{\Omega}}\equiv f(\bm{y})\bm{\Omega}$, where $f$ is some differentiable scalar field on $\mathcal{M}$, defined such that the new phase space volume element satisfies:
\begin{equation}
    \mathcal{L}_{\bm{v}}\bar{\bm{\Omega}}=\mathcal{L}_{\bm{v}}(f\bm{\Omega})=0.
\end{equation}
The condition on $f$ for such a statement to be true may be obtained by expanding the Lie derivative using the product rule, 
\begin{equation}\label{eqn:NH lie expansion}
    \mathcal{L}_{\bm{v}}(f\bm{\Omega})=\left(\bm{v}(f)+f\text{div}_{\bm{\Omega}}\bm{v}\right)\bm{\Omega}=0.
\end{equation}
For non-trivial $\bm{\Omega}$, \eref{eqn:NH lie expansion} provides a differential transport equation for $f$ through phase space 
\begin{equation}\label{eqn: differential transport}
    \bm{v}(f)+f\text{div}_{\bm{\Omega}}\bm{v}=0.
\end{equation} 

\begin{figure}[!ht]
    \centering
    \vspace{10pt}
    \includegraphics[scale=1.00]{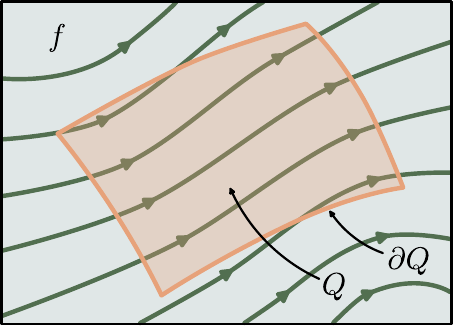}
    \vspace{5pt}
    \caption{\label{fig: phase space with volume with f}Introduction of the distribution function $f$, a scalar field on the manifold $\mathcal{M}$ along with an integration domain $Q$ with volume element $\bm{\Omega}$ and boundary $\partial Q$ with hypersurface element $\bm{\omega}$. The integral $\int_{\partial Q}f\bm{\omega}$ then measures the number of worldlines (particles) passing through the hypersurface $\partial Q$.}
\end{figure}

Further, the function $f$ may be identified by considering Stokes' theorem applied to an arbitrary region of phase space $Q\in\mathcal{M}$: 
\begin{equation} \label{eqn: integral transport}
    \int_Q\mathcal{L}_{\bm{v}}(f\bm{\Omega})=\int_Q\bm{\mathrm{d}}(f\bm{\omega})
    =\int_{\partial Q}f\bm{\omega}=0,
\end{equation}
where $\partial Q$ is the bounding surface of $Q$. The measure $\int_{\partial Q}f\bm{\omega}$ can then freely be identified with the number of worldlines (particles) passing through the closed hypersurface $\partial Q$, and $f\bm{\omega}$ the flux of particles through $\partial Q$. Hence \eref{eqn: integral transport} simply states that the divergence in phase space of the particle flux is equal to zero, i.e. the number of worldlines (particles) entering the region $Q$ is equal to the number leaving (see \fref{fig: phase space with volume with f}). Therefore $f$ may be defined as the phase space density of worldlines (particles) on hypersurfaces in $\mathcal{M}$ or, more colloquially, the particle distribution function.

\subsection{Adding Collisions}\label{subsec: adding collisions}
\begin{figure}[!ht]
    \centering
    \vspace{10pt}
    \includegraphics[scale=1.00]{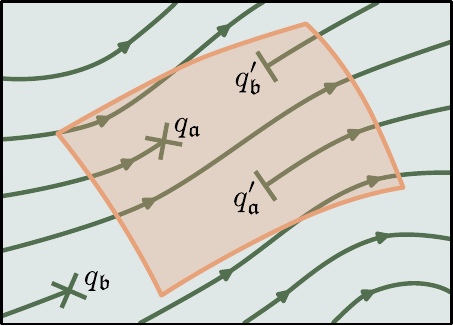}
    \vspace{5pt}
    \caption{\label{fig:adding collisions}Collisions (discrete transfers of momentum) are described by the termination (crosses) and starting (bar) of particle worldlines within phase space. Here a particle $\mathfrak{a}$ at a position $q_\mathfrak{a}$ collides with a particle $\mathfrak{b}$ at a position $q_\mathfrak{b}$, terminating each worldline. The locations $q_\mathfrak{a}$ and $q_\mathfrak{b}$ need not be coincident as the particles may have different momenta (a coordinate in phase space). The collision may cause a discrete exchange of momentum, whereby the particles then re-appear at points $q'_\mathfrak{a}$ and $q'_\mathfrak{b}$, starting a new pair of worldlines. The integral $\int_Q\bm{C}$ then counts the number of collisions (worldlines that are started and/or terminated) within the volume $Q$.}
\end{figure}

The differential and integral transport equations for the distribution function, \esref{eqn: differential transport} and \eqref{eqn: integral transport}, assume an absence of discrete, discontinuous interactions (collisions) within $\mathcal{M}$ between particles. A \textit{collision} at a point $q\in\mathcal{M}$ may be defined as a termination or starting of a worldline (more precisely the occupation of a worldline by a particle) due to a particle being destroyed/created or receiving a discrete transfer of momentum. For example, a particle $\mathfrak{a}$\footnote{Particle species are referred to by Gothic scripts in this paper.} at a position $q_\mathfrak{a}$ in its phase space $\mathcal{M}_\mathfrak{a}$ may receive a discrete quantity of momentum due to scattering off a particle $\mathfrak{b}$ at position $q_\mathfrak{b}\in\mathcal{M}_\mathfrak{b}$. The scattering causes the termination of the worldlines of particles $\mathfrak{a}$ and $\mathfrak{b}$ at points $q_\mathfrak{a}$ and $q_\mathfrak{b}$ respectively as each particle's momentum (a coordinate in phase space) has changed by a discrete amount. The worldlines for particles $\mathfrak{a}$ and $\mathfrak{b}$ are then re-started at some new positions $q'_\mathfrak{a}$ and $q'_\mathfrak{b}$ as shown in \fref{fig:adding collisions}. Alternatively, particle $\mathfrak{a}$ may undergo some emissive or decay process, where again it will undergo some discrete change in momentum and therefore terminate and re-start its worldline, while also starting new worldlines for any secondary particles emitted. Interactions, scattering, collisions, emissive and decay process all have the same effect in phase space, that being the termination and starting of worldlines, hence the term \textit{collision} will be used to refer collectively to these processes.

These collisions can then be represented by a new measure $\bm{C}(\bm{y})\propto \bm{\Omega}$ known as the collision integral, which ``counts'' the number of collisions within the volume $Q$ (see further in \sref{subsec: coll integral cases}). This measure can therefore be simply added to the right hand side of \esref{eqn: differential transport} and \eqref{eqn: integral transport}, to give the differential transport equations for $f$:
\begin{equation}\label{eqn: differential transport w coll}
    \bm{v}(f)+f\text{div}_{\bm{\Omega}}\bm{v}=\bm{C},
\end{equation}
and its integral form:
\begin{equation}\label{eqn: integral transport w coll}
    \int_{\partial Q}f\bm{\omega}=\int_Q \bm{C}.
\end{equation}

%% file: Sections/BE_coordinate.tex
\section{Coordinate-Dependent Transport Equation}\label{sec:coordinate}

\subsection{Phase Flow with Coordinates}\label{subsec: phase flow w coords}

Though \esref{eqn: differential transport w coll} and \eqref{eqn: integral transport w coll}, are mathematically elegant in their coordinate-free form, some definition of coordinates (basis vectors) are needed to give them utility in modelling astrophysical sources.

For this purpose, two sets of basis vectors are used. The nature of spacetime is assumed to be a 4D Riemannian manifold $\mathcal{N}$, with signature $(-+++)$, and a coordinate basis $\bm{e}_\alpha=$ (Greek scripts) with coordinates $x^\alpha=\{x^0,...,x^3\}$\footnote{The choice of a precise set of coordinates, e.g. Cartesian, cylindrical, spherical, Schwarzschild, Boyer-Lindquist, etc. remains arbitrary. As such this model can be applied to a wide range of spacetimes.}. For momentum space, which exists in the tangent space $T_\mathcal{N}$ to $\mathcal{N}$, a local-orthonormal basis (tetrad) $\bm{e}_a$ (Latin scripts), with components $p^a=\{p^0,...,p^3\}$ is used. The worldline of a particle through the 8-dimensional phase space is then given by 
\begin{equation}\label{eqn: worldline}
    \frac{\mathrm{d}x^\alpha}{\mathrm{d}\lambda}=p^a e^\alpha_a, \quad \frac{\mathrm{d}p^a}{\mathrm{d}\lambda}=-\Gamma^a_{~bc}p^bp^c+mF^a,
\end{equation}
where the transformation between the two basis is $\bm{e}_a=e_a^{~\alpha}\bm{e}_\alpha$, $\Gamma^a_{~bc}$ are the connection coefficients in the local orthonormal basis (i.e. Ricci rotation coefficients) and $F^a$ are the components of some arbitrary four-force $\bm{F}$. For massive particles $m\neq 0$ the affine parameter $\lambda$ is related to the proper time $\tau$ by $m\lambda=\tau$. The manifold $\mathcal{M}$, corresponding to the 8-dimensional phase space, is given by the set 
\begin{equation}
\begin{split}
    \mathcal{M}=&\{(\bm{x},\bm{p}):\bm{x}\in \mathcal{N},\bm{p}\in\bm{T}_\mathcal{N},
    \\
    &\quad\bm{p}^2\leq0,\bm{p}\text{ future directed}\}.
    \end{split}
\end{equation}
The phase flow in $\mathcal{M}$ is generated by the Lie derivative with respect to $\bm{v}$, where, using \eref{eqn: worldline} 
\begin{equation}\label{eqn:phase flow vector}
    \bm{v}=p^ae_a^{~\alpha}\frac{\partial}{\partial x^\alpha}+\left(-\Gamma^a_{~bc}p^bp^c+mF^a\right)\frac{\partial}{\partial p^a}.
\end{equation}
The rest mass of a particle \begin{equation}        
    -m^2=\bm{p}^2=g_{\alpha\beta}e^\alpha_{~a}e^\beta_{~b}p^ap^b,
\end{equation}
is a scalar function on $\mathcal{M}$ which, on physicality grounds, should be a constant with respect to the phase flow i.e. 
\begin{equation}
    \mathcal{L}_{\bm{v}}(-m^2)=0.
\end{equation}
Via direct calculation\footnote{This requires the additional relation between the connection coefficients in a orthonormal basis $\Gamma^a_{~bc}$ (Ricci rotation coefficients) and those in a coordinate basis $\Gamma^{\alpha}_{~\beta\gamma}$ (Christoffel symbols): 
\begin{equation}
    \Gamma^a_{~bc}=e^a_{~\alpha}e_c^{~\gamma}\left(\partial_\gamma e_b^{~\alpha}+\Gamma^\alpha_{~\beta\gamma}e_b^{~\beta}\right).
\end{equation}}, this yields the following constraint on the force $\bm{F}$: 
\begin{equation}
    \bm{F}\cdot\bm{p}=0,
\end{equation}
which is met for all forces expected to occur within an astrophysical setting (e.g. Lorentz and radiation reaction forces). With $m^2$ being constant with respect to the phase flow, particles exist only on hypersurfaces in $\mathcal{M}$ defined by $m^2=\text{const}$. This restricted manifold is now denoted $\mathcal{M}_m$ and is 7-dimensional with coordinates $x^\alpha=\{x^0,...,x^3\}$ and $p^i=\{p^1,...,p^3\}$, where second-half Latin alphabet letters refer to indices valued $i\in\{1,2,3\}$, where $p^0$, the ``energy'' of the particle, has been chosen to be the dependent variable defined by 
\begin{equation}
    p^0=\sqrt{m^2+(p^i)^2}.
\end{equation}
The phase flow on the restricted manifold $\mathcal{M}_m$ is then given by
\begin{equation}\label{eqn: restricted phase flow}
    \bm{v}=p^ae_a^{~\alpha}\frac{\partial}{\partial x^\alpha}+\left(-\Gamma^i_{~bc}p^bp^c+mF^i\right)\frac{\partial}{\partial p^i}.
\end{equation}

What are the volume forms on $\mathcal{M}_m$? The spacetime volume form is given by the standard form\footnote{The Hodge dual of the unit function $\bm{\chi}=\star 1$.} 
\begin{equation}\label{eqn:x space volume form in coordinates}
    \bm{\chi}=\frac{\chi_{\alpha\beta\gamma\delta}(\bm{x})}{4!}\bm{\mathrm{d}}x^{\alpha\beta\gamma\delta},
\end{equation}
where $\chi_{\alpha\beta\gamma\delta}$ are components of the completely antisymmetric Levi-Civita tensor, with $\chi_{1234}=\sqrt{-\det(g_{\alpha\beta})}=\sqrt{-g}$. The momentum space volume, restricted to $\mathcal{M}_m$, is likewise given by 
\begin{equation}
    \bm{\pi}=2\Theta(p^0)\delta(\bm{p}^2-m^2)\sqrt{-\eta}\bm{\mathrm{d}}p^{0123}=\frac{\sqrt{-\eta}}{p^0}\bm{\mathrm{d}}p^{123},
\end{equation}
where $\eta_{ab}$ is the metric in momentum space. For an orthonormal basis $\sqrt{-\eta}=\sqrt{-\det(\eta_{ab})}=1$, hence 
\begin{equation}\label{eqn:p space volume form in coordinates}
    \bm{\pi}=\frac{1}{p^0}\bm{\mathrm{d}}p^{123}=\frac{\pi_{ijk}}{3!}\bm{\mathrm{d}}p^{ijk}.
\end{equation}
The total volume form is then given by the wedge product of \esref{eqn:x space volume form in coordinates} and \eqref{eqn:p space volume form in coordinates}: 
\begin{equation}\label{eqn:volume form}
    \bm{\Omega}=\bm{\chi}\wedge\bm{\pi}.
\end{equation}
From \eref{eqn:volume form}, the hypersurface element $\boldsymbol{\omega}$ orthogonal to the phase flow $\bm{v}$ (\eref{eqn: restricted phase flow}) in $\mathcal{M}_m$ can also be evaluated: 
\begin{equation}
\begin{split}\label{eqn:surface forms in coordinates non-observer}
    \bm{\omega}\equiv\bm{v}\lrcorner\bm{\Omega}=&p^ae_a^{~\alpha}\frac{\chi_{\alpha\beta\gamma\delta}(\bm{x})}{3!}\bm{\mathrm{d}}x^{\beta\gamma\delta}\wedge\bm{\pi}
    \\
    &+\bm{\chi}\wedge\left(-\Gamma^i_{~bc}p^bp^c+mF^i\right)\frac{\pi_{ijk}}{2!}\bm{\mathrm{d}}p^{jk}.
\end{split}
\end{equation}
As discussed in \sref{subsec: distribution function}, for a system to obey Liouville's theorem $\bm{\mathrm{d}}\bm{\omega}=0$ is required, by consideration of \eref{eqn:surface forms in coordinates non-observer} this condition is equivalent to $\frac{\partial F^a}{\partial p^a}=0$. As an example where this condition is met, the Lorentz force on a particle of charge $q$ and mass $m$ is given by $F^a=\frac{q}{m} F^{ab}p_b$, where $F^{ab}$ is the electromagnetic field tensor, therefore $\frac{\partial F^a}{\partial p^a}=F^a_{~a}=0$ as $F^{ab}$ is antisymmetric. However, a counter example is the radiation reaction force\footnote{The radiation reaction force is also known as the Abraham-Lorentz-Dirac force \citep{Abraham_1905,Lorentz_1892,Dirac_1938} \citep[see also][for more modern descriptions]{GrallaEtAl_2009,PoissonEtAl_2011}.} experienced by a charged particle being accelerated by an electromagnetic field where $F^a\propto\left(\eta^{ab}+\frac{p^a p^b}{m^2c^2}\right)\left(\frac{q}{m}p^dp^c\partial_cF_{bd}+\frac{q^2}{m^2}qF_{bc}F^{cd}p_d\right)$, for such a force it can be shown that $\frac{\partial F^a}{\partial p^a}\neq0$, and is therefore a non-conservative force, evident by the fact that it results in the charged particle losing energy over time.

As such non-conservative forces, e.g. radiation reaction, are a necessary inclusion of any self-consistent astrophysical model; hence $\mathbf{d}\bm{\omega}\neq0$ in general and the correct forms for the transport equation for the particle distribution function are \esref{eqn: differential transport w coll} and \eqref{eqn: integral transport w coll}.


\subsection{Stationary Observers}\label{sec: stationary observers}
It is important to define a set of observers with respect to whom the distribution function and its derived properties are measured. \textit{If the spacetime is stationary}, it permits a set of Eulerian observers. Their worldlines are defined everywhere by the timelike unit vector $\bm{n}=n_\alpha\bm{e}^\alpha=n_0\bm{\mathrm{d}}t$, with $\bm{n}\cdot\bm{n}=n^0n_0=-1$. At each value of proper time $\tau$ along the worldline of an observer, there exists a unique, spacelike hypersurface $\Sigma_\tau$ orthogonal to the worldline. This hypersurface defines the set of events that are considered simultaneous with respect to that observer at that value of proper time.

\begin{figure}[!ht]
    \centering
    \includegraphics[scale=1.00]{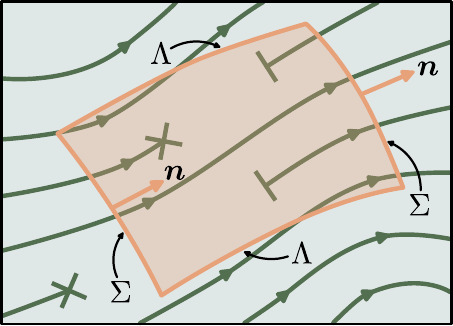}
    \vspace{5pt}
    \caption{\label{fig: phase space with observer}Timelike $\Sigma$ and spacelike $\Lambda$ hypersurfaces in phase space $\mathcal{M}_m$ defined by a set of stationary observers with trajectory tangent to $\bm{n}$. The spacelike hypersurfaces $\Sigma$ define a set of simultaneous events as measured by the observer.}
\end{figure}

The hypersurface element $\bm{\omega}$, given by \eref{eqn:surface forms in coordinates non-observer}, may be decomposed into hypersurface elements of simultaneous and non-simultaneous events as measured by a stationary observer: 
\begin{equation}\label{eqn:surface forms in coordinates observer}
    \begin{split}
    \bm{\omega}&=\left[(-p^an_a)n^\alpha\frac{\chi_{\alpha\beta\gamma\delta}(\bm{x})}{3!}\bm{\mathrm{d}}x^{\beta\gamma\delta}\wedge\bm{\pi}\right]
    \\
    &\quad+\left[\left(p^ae^{~\alpha}_a+p^an_a n^\alpha\right)\frac{\chi_{\alpha\beta\gamma\delta}(\bm{x})}{3!}\bm{\mathrm{d}}x^{\beta\gamma\delta}\wedge\bm{\pi}\right.
    \\
    &\quad+\left.\bm{\chi}\wedge\left(-\Gamma^i_{~bc}p^bp^c+mF^i\right)\frac{\pi_{ijk}}{2!}\bm{\mathrm{d}}p^{jk}\right]
    \\
    &=\bm{\omega}_{\Sigma}+\bm{\omega}_{\Lambda}.
    \end{split}
\end{equation}
The first square bracketed term in \eref{eqn:surface forms in coordinates observer} is $\bm{\omega}_{\Sigma}$, referring to the element of $\Sigma$ hypersurfaces, while the second square bracket is the element of timelike hypersurfaces $\Lambda$ of non-simultaneous events, $\bm{\omega}_{\Lambda}$ (see \fref{fig: phase space with observer}).

With a set of observers, the total number of particles in the system can be counted in a well defined way. Consider the insight from \sref{subsec: distribution function}, that the $\int_{\partial Q} f\bm{\omega}$ is the number of worldlines passing through some hypersurface defined by the boundary $\partial Q$. Therefore, taking $\partial Q$ to be the open hypersurface $\Sigma$, $\int_{\Sigma}f\bm{\omega}=N$ the number of particles measured simultaneously passing through that surface, i.e. the total number of particles within the manifold at a given time $\tau$ according to that local observer. If $\Sigma=X\times P$ is decomposed into the product of a spatial submanifold $X$ and the momentum-space submanifold $P$ this integral can be expanded as:
\begin{equation}\label{eqn: particle number}
\begin{split}
    N&=\int_{\Sigma}f\bm{\omega}_{\Sigma}
    \\
    &=\int_X -n_a n^\alpha\frac{\chi_{\alpha\beta\gamma\delta(\bm{x})}}{3!}\bm{\mathrm{d}}x^{\beta\gamma\delta}\int_{P}fp^a\bm{\pi}.
\end{split}
\end{equation}
The latter part of \eref{eqn: particle number} can be identified as the first moment of the distribution function i.e. locally measured 4-flow vector: 
\begin{equation}\label{eqn: 4-flow}
    N^a=\int_{P}fp^a\bm{\pi},
\end{equation}
from which the scalar number density of particles $n$ as measured by a stationary observer can be defined:
\begin{equation}\label{eqn: scalar number density}
    n=-n_aN^a.
\end{equation}
Higher moments of the distribution function can similarly be defined, for example the locally measured stress-energy tensor is defined by 
\begin{equation}\label{eqn: stress energy tensor}
    T^{ab}=\int_{P}fp^ap^b\bm{\pi},
\end{equation}
through which the scalar energy density $e=n_an_bT^{ab}$ and pressure $p= \frac{1}{3}T^{ab}\Delta_{ab}$ (where $\Delta_{ab}=\eta_{ab}+n_an_b$ is the projection tensor) can be defined. 

An alternative to the set of Eulerian observers is a set of observers co-moving with some bulk flow. However, for co-moving observers, their local tetrad $\boldsymbol{e}_a$ is a function of the bulk flow, and therefore variable, unlike a Eulerian observer whose tetrad is fixed. This results in the hypersurface elements \eref{eqn:surface forms in coordinates non-observer} also becoming a function of the flow and therefore so too would the numerical fluxes (see \sref{subsec: DIP transport}) used to evaluate transport of the distribution function, thereby requiring unnecessary recalculation when the bulk flow changes.


\subsection{The Collision Integral}\label{subsec: coll integral cases}

The collision integral $\bm{C}(\bm{x},\bm{p})$ is a measure of the number of worldlines that are terminated and started within some volume of phase space. It is useful to relate this to the transition probability $w_{\mathfrak{if}}$ for an initial state $\mathfrak{i}$ of particles to transition to a final state $\mathfrak{f}$. Modifying the results of \citet{BerestetskiiEtAl_1982}, the differential transition probability may be written:
\begin{equation}\label{eqn: diff transition}
    \bm{w}_\mathfrak{if}=\delta^{(4)}(\bm{p}_\mathfrak{i}-\bm{p}_\mathfrak{f})\left|T_\mathfrak{if}\right|^2\bm{\chi},
\end{equation}
where the Dirac delta function describes conservation of momentum during the transition and $T_\mathfrak{if}$ is the standard scattering $T$-matrix.

Given a distribution of particles $f(\bm{x},\bm{p})$, the number of particle worldlines crossing a $\Sigma$ hypersurface is given by \eref{eqn: particle number}. Assuming a well defined basis transformation $n_a=(-1,0,0,0)$ and \eref{eqn: particle number} may be integrated over a test region in the spatial sub-manifold $X$, giving
\begin{equation}
    \int_{X}\bm{\omega}_{\Sigma}=p^0V\bm{\pi},
\end{equation}
where $V$ is some spatial volume. As such the expression $p^0V\bm{\pi}$ can be thought of as the momentum-space number density of particles with a unity distribution function, i.e. 1 particle per unit hypersurface element. The differential transition probability, \eref{eqn: diff transition}, may then be expanded to include the incident and final particle's momentum states:
\begin{equation}
    \bm{w}_\mathfrak{if}=\delta^{(4)}(\bm{p}_\mathfrak{i}-\bm{p}_\mathfrak{f})\frac{\left|T_\mathfrak{if}\right|^2}{\mathfrak{n}_\mathfrak{i}!\mathfrak{n}_\mathfrak{f}!}\bm{\chi}\prod_{\mathfrak{a}\in\mathfrak{i}}p^0_\mathfrak{a}V\bm{\pi}_{\mathfrak{a}}\prod_{\mathfrak{b}\in\mathfrak{f}}p^0_\mathfrak{b}V\bm{\pi}_{\mathfrak{b}},
\end{equation}
where $\mathfrak{n}_\mathfrak{i}$ and $\mathfrak{n}_\mathfrak{f}$ are the number of identical initial and identical final states respectively, which are included to avoid over counting states.

In the literature, the factors of $p^0V$ are often absorbed into the scattering matrix $T_\mathfrak{if}$ to generate the more familiar scattering matrix $M_\mathfrak{if}$ i.e.
\begin{equation}\label{eqn: T to M scattering}
    M_\mathfrak{if}=T_\mathfrak{if}\sqrt{\prod_{\mathfrak{a}\in\mathfrak{i}}p^0_\mathfrak{a}V\prod_{\mathfrak{b}\in\mathfrak{f}}p^0_\mathfrak{b}V},
\end{equation}    
giving
\begin{equation}
    \bm{w}_\mathfrak{if}=\delta^{(4)}(\bm{p}_\mathfrak{i}-\bm{p}_\mathfrak{f})\frac{\left|M_\mathfrak{if}\right|^2}{\mathfrak{n}_\mathfrak{i}!\mathfrak{n}_\mathfrak{f}!}\bm{\chi}\prod_{\mathfrak{a}\in\mathfrak{i}}\bm{\pi}_{\mathfrak{a}}\prod_{\mathfrak{b}\in\mathfrak{f}}\bm{\pi}_{\mathfrak{b}}.
\end{equation}

The number of worldlines terminated or started by an collision $\mathfrak{i}\rightarrow \mathfrak{f}$ is then given by reintroducing non-unity distribution functions for the initial particles:
\begin{equation}\label{eqn: N worldlines}
    \bm{N}_\mathfrak{if}=\delta^{(4)}(\bm{p}_\mathfrak{i}-\bm{p}_\mathfrak{f})\frac{\left|M_\mathfrak{if}\right|^2}{\mathfrak{n}_\mathfrak{i}!\mathfrak{n}_\mathfrak{f}!}\bm{\chi}\prod_{\mathfrak{a}\in\mathfrak{i}}f_\mathfrak{a}(\bm{x},\bm{p})\bm{\pi}_{\mathfrak{a}}\prod_{\mathfrak{b}\in\mathfrak{f}}\bm{\pi}_{\mathfrak{b}}.
\end{equation}
With the volume element for phase space given by \eref{eqn:volume form}, it can be recognised that the quantity  $\bm{N}_\mathfrak{if}$ is proportional to $\bm{\Omega}$, and therefore related to the collision integral $\bm{C}(\bm{x},\bm{p})$. To understand this relation, consider a particle of type-$\mathfrak{a}$. The number of its worldlines terminated by collisions where it is a member of the incident state $\mathfrak{i}$ is given by $\sum_{\mathfrak{a}\in\mathfrak{i}}\bm{N}_\mathfrak{if}$, and similarly the number of its worldlines that are started is $\sum_{\mathfrak{a}\in\mathfrak{f}}\bm{N}_\mathfrak{if}$.

The nature of the collision integral for a single particle of type-$\mathfrak{a}$ is therefore
\begin{equation}\label{eqn: diff collision integral}
\bm{C}(\bm{x},\bm{p}_\mathfrak{a})=\sum_{\mathfrak{a}\in\mathfrak{f}}\bm{N}_\mathfrak{if}-\sum_{\mathfrak{a}\in\mathfrak{i}}\bm{N}_\mathfrak{if}.
\end{equation}
Specific examples of the collision integral for binary (scattering) collisions i.e. $\mathfrak{ab}\rightarrow\mathfrak{cd}$ and emission (decay) processes i.e. $\mathfrak{a}\rightarrow\mathfrak{bc}$ are given in \aref{app: abs and emi}.

\eref{eqn: diff collision integral} contains integrals over the momentum states of all the initial and final state particles; it is therefore of high dimensionality and further complicated by potentially non-linear dependence on the particle distribution functions. As such, it is generally impractical to use in this form unless some methodology for simplifying the integral is applied, or some assumptions are made for the symmetries of $f$ in momentum space---traditionally, isotropy.

%% file: Sections/DIP.tex
\section{Distribution-In-Plateaux}\label{sec:DIP}

\begin{figure*}
    \centering
    \includegraphics[scale=1.00]{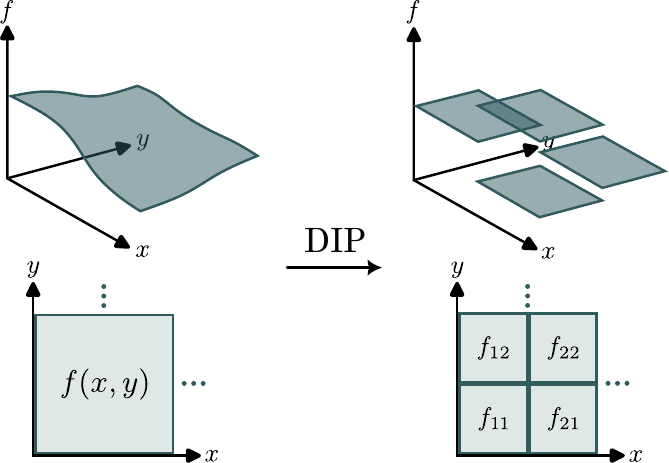}
    \vspace{5pt}
    \caption{\label{fig: DIP}The continuous distribution function $f$ for each particle is discretised via \eref{eqn: DIP} to form a range of plateaux over phase space (here taken to be two dimensional with coordinates $x$ and $y$) indexed by the values $f_{ij}$, hence the name Distribution-In-Plateaux.}
\end{figure*}

The integrated form of the Boltzmann equations (\eref{eqn: integral transport w coll}) is mathematically elegant but in its present form is difficult to evaluate. The difficulty arises from the integrals being evaluated over the entire volume of phase space. This can be simplified by splitting the integration domain $Q$ into a set of continuous sub-domains with common boundaries. Therefore the left hand side of \eref{eqn: integral transport w coll} describes the flux of particle worldlines through the boundaries of each sub-domain, while the right hand side counts collisions within each sub-domain. As all boundaries are shared between sub-domains, the number of particle worldlines is conserved over the whole domain.

To aid in simplifying the equations in this section, the set of spacetime coordinates will be denoted by $x^\alpha=\{x^0,x^1,x^2,x^3\}=\{t,x,y,z\}$, making the time coordinate $t$ explicit, and spatial coordinates $\{x,y,z\}$ refer, abstractly, to any set of three chosen spatial coordinates not just the standard Cartesian set i.e. $\{x,y,z\}$ (see \aref{app:metrics} for examples). For momentum-space, coordinates are concretely defined to be $p^i=\{p^1,p^2,p^3\}=\{p,u,\phi\}$ the set of modified spherical coordinates ($u=\cos\theta$), as this is the set used for the numerical evaluation of collision integrals in \citetalias{PaperII}. 

With these coordinate sets, the procedure for splitting the integration domain $Q$ into a continuous set of sub-domains can formally be achieved by taking the particle distribution function to be constant within each sub-domain of phase space. The total distribution is formed by the sum of these constant regions: 
\begin{equation}
\begin{split} \label{eqn: DIP}
    f(\bm{x},\bm{p}) =& \sum_{\alpha\beta\gamma\delta ijk} \frac{f_{\alpha\beta\gamma\delta ijk}}{p^2\Delta p_i\Delta u_j\Delta\phi_k} H_{\alpha}(t)H_{\beta}(x)
    \\
    &\times H_{\gamma}(y)H_{\delta}(z)H_{i}(p)H_{j}(u)H_{k}(\phi),
\end{split}
\end{equation}
where indices $\alpha,\beta,\gamma,\delta,i,j,$ and $k$ now refer to a discrete set of phase-space coordinates and $H$ is a boxcar function. As an example, for the $t$ coordinate the function $H$ is defined as
\begin{align} \label{eqn: DIP Box Example}
\begin{split}
    H_{\alpha}(t)&=\Theta(t-t_\alpha) - \Theta(t-t_{\alpha+1})
    \\
    &=
    \begin{cases}
        1, & \text{for } t_{\alpha}<t<t_{\alpha+1},
        \\
        h_\alpha^- & \text{for } t=t_{\alpha} 
        \\
        h_\alpha^+=1-h_\alpha^- & \text{for } t=t_{\alpha+1} 
        \\
        0, & \text{for all other } t,
    \end{cases}
\end{split}
\end{align}
with $\Theta$ being the Heaviside step function and $h_\alpha^\pm$ being the boxcar function's values at the bounds of the domain $t_\alpha\le t\le t_{\alpha+1}$\footnote{By allowing the definition of $\Theta(0)$ to be a variable between 0 and 1, usually taken to be 0,1 or $1/2$, modification of the numerical scheme may be made, as described in \citetalias{PaperII}, while maintaining conservative particle transport.}. 

Through the action of \eref{eqn: DIP}, the continuous distribution function is replaced by a series of plateaux, each with a ``height'' of $f_{\alpha\beta\gamma\delta ijk}$ over a discrete sub-domain in phase space (see \fref{fig: DIP}), and hence termed \textit{Distribution-In-Plateaux} (DIP)\footnote{It has been pointed out to the authors that this method of discretisation may be mapped to a zeroth-order discontinuous Galerkin method, which has been developed independently of this work, and further, has been shown to have good conservative properties for number density and energy \citep[see, for example][]{CockburnShu_2001,GambaRjasanow_2018,GasparyanEtAl_2022}.}. This nomenclature has similarities with that of the Particle-In-Cell (PIC), whereby the distribution function is considered as a sum over Dirac delta functions.

The factor of $p^2\Delta p_i\Delta u_j\Delta\phi_k$ in \eref{eqn: DIP} is not strictly required, rather it has been introduced for convenience such that according to \eref{eqn: scalar number density} the scalar number density $n$ of particles measured by the stationary observer, within a spatial sub-domain $X_{\beta\gamma\delta}$ and between times $t_\alpha$ and $t_{\alpha+1}$, is given by
\begin{equation}
    n=\sum_{ijk}f_{\alpha\beta\gamma\delta ijk},
\end{equation}
and as such $f_{\alpha\beta\gamma\delta ijk}$ can be recognised as the number density of particles within a momentum space sub-domain $P_{ijk}$.

The application of DIP form of the distribution function to the equations of phase-space transport and collision integrals simplifies their evaluation in terms of expressions that can be implemented computationally. However, this comes at the cost of these expressions becoming rather long. These long expressions can be found in full in Appendices \ref{app: abs and emi} and \ref{app:longeq}, which will be referred to in the following sections.


\subsection{DIP Applied to Collision Integrals}\label{subsec: DIP coll int}

Before describing how DIP simplifies the larger problem of phase-space transport, it is useful to consider how it simplifies the problem of evaluating collision integrals. For a binary collision $\mathfrak{ab\rightarrow cd}$, the gain of particles of type-$\mathfrak{c}$ is given by the collision integral (\eref{eqn: diff collision integral}):
\begin{equation}
\begin{split}\label{eqn: collision integral example for DIP}
    \int \bm{C}(\bm{x},\bm{p}_\mathfrak{c})=&\int f_{\mathfrak{a}}(\bm{x},\bm{p}_\mathfrak{a})f_{\mathfrak{b}}(\bm{x},\bm{p}_\mathfrak{b})
    \\ 
    &\times G_{\mathfrak{a}\mathfrak{b}\rightarrow\mathfrak{c}\mathfrak{d}}\mathrm{d}^3p_\mathfrak{a}\mathrm{d}^3p_\mathfrak{b}\mathrm{d}^3p_\mathfrak{c} \bm{\chi},
\end{split}
\end{equation}
where $G_{12\rightarrow34}$ is the ``gain term'' given by \eref{eqn: binary gain term} in \aref{app: abs and emi}. The integration domain for this term is the phase space of particle $\mathfrak{c}$ and the momentum space of particles $\mathfrak{a}$ and $\mathfrak{b}$. By deploying DIP, the value of the distribution functions of particles $\mathfrak{a},\mathfrak{b}$, and $\mathfrak{c}$ become constant within a sub-domains of phase space and, therefore, can be taken outside integrals over those sub-domains. Consider a momentum-space sub-domain for particle $\mathfrak{c}$ to be $P_{ijk}=[p_{i},p_{i+1}]\times[u_{j},u_{j+1}]\times[\phi_{k},\phi_{k+1}]$, with DIP deployed, the collision integral (\eref{eqn: collision integral example for DIP}) over this sub-domain becomes:   
\begin{equation}
\begin{split}
    \int^{t_{\alpha+1}}_{t_{\alpha}}\int_{X_{\beta\gamma\delta}}&\int_{P_{ijk}} \bm{C}(\bm{x},\bm{p}_\mathfrak{c}) = \sum_{lmnopq}f_{\mathfrak{a},\alpha\beta\gamma\delta lmn}
    \\
    &\times f_{\mathfrak{b},\alpha\beta\gamma\delta  opq}G_{\mathfrak{a}\mathfrak{b}\rightarrow\mathfrak{c}\mathfrak{d},ijklmnopq}\mathcal{V}_{\alpha\beta\gamma\delta},
\end{split}
\end{equation}
where the gain term has become a 9-dimensional ``gain array'' $G_{\mathfrak{a}\mathfrak{b}\rightarrow\mathfrak{c}\mathfrak{d},ijklmnopq}$ given by \eref{eqn: binary gain matrix}, and $\mathcal{V}_{\alpha\beta\gamma\delta}$ is the spacetime volume element given by \eref{eqn: DIP volume element} found in \aref{app:longeq}. The elements of the gain array are integrals over different momentum-space sub-domains of the incoming and outgoing particles, which, being independent of the distribution of those particles, are pre-computable by numerical means (see \citetalias{PaperII}), and are proportional to the rate of a collision. Further, elements of the gain array are independent of spatial coordinates, due to the use of a local-orthonormal basis for momentum space, and as such are valid at all points in spacetime, independent of the spacetime's nature. This process of discretisation and integration can be applied to all collision terms equally, thereby all collisions are presented as pre-computed arrays within the DIPLODOCUS framework. 

The application of DIP to collision integrals presents some immediate advantages of using DIP over PIC. As all collision array terms may be pre-evaluated and reused over multiple simulations of phase-space transport, there is a reduction in computational intensity compared to the calculation of collisions during transport---especially when the number of particles (collisions) is large. In addition, it provides uniform sampling of collisions over the entire phase-space domain---independent of the number of particles in the system.


\subsection{DIP Applied to Transport}\label{subsec: DIP transport}
Consider an arbitrary but stationary spacetime, with a metric tensor given by:
\begin{equation}
\begin{split}\label{eqn: stationary metric general}
\bm{g} =& \left(-A^2+\frac{E^2}{B^2}+\frac{F^2}{C^2}+\frac{G^2}{D^2}\right)\bm{\mathrm{d}}t\otimes\bm{\mathrm{d}}t
\\
&+B^2\bm{\mathrm{d}}x\otimes\bm{\mathrm{d}}x+C^2\bm{\mathrm{d}}y\otimes\bm{\mathrm{d}}y+D^2\bm{\mathrm{d}}z\otimes\bm{\mathrm{d}}z
\\
&+2E\bm{\mathrm{d}}t\otimes\bm{\mathrm{d}}x+ 2F\bm{\mathrm{d}}t\otimes\bm{\mathrm{d}}y+ 2G\bm{\mathrm{d}}t\otimes\bm{\mathrm{d}}z,
\end{split}
\end{equation}
Being stationary, the functions $A,B,C,D,E,F,$ and $G$ are functions of a subset of the spatial coordinates $x^\iota$ only. The functions $E,F,G$ form what is conventionally known as the \textit{shift vector} $\beta_\iota=(E,F,G)$ with the function $A$ being the \textit{lapse function}. The metric given by \eref{eqn: stationary metric general} could refer to a wide range of metrics commonly applied to astrophysical settings, an example subset of which are given in \aref{app:metrics}.

\begin{figure*}[!ht]
    \centering
    \includegraphics[scale=1.00]{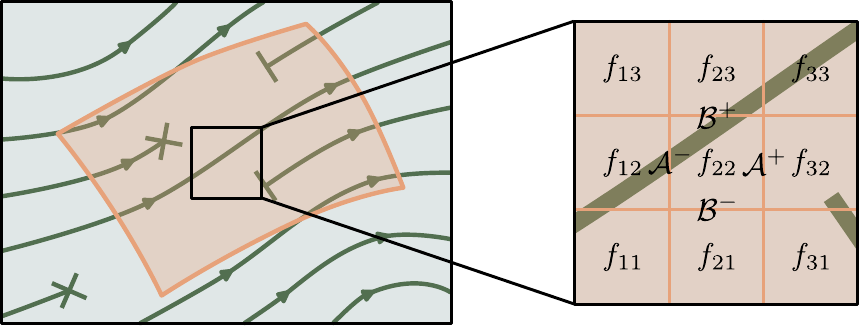}
    \vspace{5pt}
    \caption{\label{fig: simplified DIP transport}Simplified schematic of how DIP applies to phase-space transport (here phase space has been reduced to two dimensions). The action of Distribution-In-Plateaux \eref{eqn: DIP}, is to discretise phase space into sub-domains in which the distribution function $f$ takes a single value ($f_{ij}$ here). The left hand side of the transport equation \eref{eqn: integral transport w coll} $\int_{\partial Q}f\bm{\omega}$ is then an integral over the boundaries of these sub-domains, which generates the fluxes between sub-domains ($\mathcal{A}$ and $\mathcal{B}$ here). For this two dimensional example, the transport equation for $f_{22}$ takes the form $\int_{\partial Q_{22}}f\bm{\omega}=\left(h^-_{32}f_{32}+h^+_{22}f_{22}\right)\mathcal{A}^+ + \left(h^-_{22}f_{22}+h^+_{12}f_{12}\right)\mathcal{A}^- + \left(h^-_{23}f_{23}+h^+_{22}f_{22}\right)\mathcal{B}^+ + \left(h^-_{22}f_{22}+h^+_{21}f_{21}\right)\mathcal{B}^-$, where the $h$ terms (\eref{eqn: DIP Box Example}) dictate the value of the distribution function on sub-domain boundaries.}
\end{figure*}

The spacetime volume element (\eref{eqn: stationary metric general}) of the metric given by \eref{eqn: stationary metric general} has the simple form:
\begin{equation}\label{eqn:volume element stationary}
    \chi_{0123}=ABCD,
\end{equation}
and a natural coordinate transform to the tetrad of a stationary (Eulerian) observer $\bm{n}=-A\bm{\mathrm{d}}t$, could be written:
\begin{equation}
\begin{split}\label{eqn:tetrad stationary}
    e_a^{~\alpha}&=\left(n^\alpha,X^\alpha,Y^\alpha,Z^\alpha\right),
    \\
    &= \begin{pmatrix}
    \frac{1}{A} & -\frac{E}{AB^2} & -\frac{F}{AC^2} & -\frac{G}{AD^2} 
    \\
    0 & \frac{X_1}{B} & \frac{X_2}{C} & \frac{X_3}{D} 
    \\
    0 & \frac{Y_1}{B} & \frac{Y_2}{C} & \frac{Y_3}{D} 
    \\
    0 & \frac{Z_1}{B} & \frac{Z_2}{C} & \frac{Z_3}{D}
    \end{pmatrix},
\end{split}
\end{equation}
where $X^\alpha=(0,\frac{X_1}{B},\frac{X_2}{C},\frac{X_3}{D})$, $Y^\alpha=(0,\frac{Y_1}{B},\frac{Y_2}{C},\frac{Y_3}{D})$, and $X^\alpha=(0,\frac{Z_1}{B},\frac{Z_2}{C},\frac{Z_3}{D})$ are a set of mutually orthonormal vectors, which are also orthogonal to $\bm{n}$. It is often standard to align $\bm{X},\bm{Y},\text{ and }\bm{Z}$ to the $\{x,y,z\}$ coordinate directions, however, they have here been left general as it can be useful to align them to some other set of vectors e.g. the local magnetic and electric field directions \citep{ChaelEtAl_2023,GellesEtAl_2025,TsunetoeEtAl_2025}. 

With $\bm{n}$ and $e_a^{~\alpha}$ defined, the integral form of the transport equation \eref{eqn: integral transport w coll} can be evaluated using DIP over a phase space sub-domain $X_{\alpha\beta\gamma\delta}\times P_{ijk}=[t_\alpha,t_{\alpha+1}]\times[x_\beta,x_{\beta+1}]\times[y_\gamma,y_{\gamma+1}]\times[z_\delta,z_{\delta+1}]\times[p_{i},p_{i+1}]\times[u_{j},u_{j+1}]\times[\phi_{k},\phi_{k+1}]$ to give \esref{eqn: DIP volume element} to \eqref{eqn: IJK m flux} (found in \aref{app:longeq}), whereby terms on the left hand side of \eref{eqn: integral transport w coll} are converted into fluxes at the boundaries of the sub-domain (see \fref{fig: simplified DIP transport}) and those on the right hand side have been converted into a series of collision arrays. These equations form a numerical evolution scheme for all particle distribution functions, including external forces and collisions between particles. Such a scheme is conservative in terms of the total number of particles in the system (in the absence of emissive processes which are free to add or subtract particles from the system), as fluxes are always balanced between sub-domains, i.e. any particles leaving one sub-domain of phase space enter another.

As an example of how these fluxes are generated, consider the specific case of Schwarzchild geometry with coordinates $x^\alpha=\{t,r,\theta,\psi\}$ (see \aref{app:metrics}, \eref{eqn: schw coordinates}). In this geometry $\bm{n}=-\sqrt{1-\frac{r_s}{r}}\bm{\mathrm{d}}t$, where $r_s$ is the Schwarzchild radius, and the coordinate transformation \eref{eqn:tetrad stationary} may be given by 
\begin{equation}\label{eqn:tetrad schwarzchild}
    e_a^{~\alpha} = \begin{pmatrix}
    \frac{1}{\sqrt{1-\frac{r_s}{r}}} & 0 & 0 & 0 
    \\
    0 & \sqrt{1-\frac{r_s}{r}} & 0 & 0 
    \\
    0 & 0 & \frac{1}{r} & 0 
    \\
    0 & 0 & 0 & \frac{1}{r\sin\theta}
    \end{pmatrix},
\end{equation}
assuming the orthonormal basis vectors $\boldsymbol{X},\bm{Y}$, and $\bm{Z}$ are aligned to the global coordinate directions $\bm{e}_r,\bm{e}_\theta$, and $\bm{e}_\phi$.

Using these definitions for the Eulerian observer $\bm{n}$ and coordinate transformations $e_a^{~\alpha}$, the fluxes through phase space sub-domain boundaries and the spacetime volume element (\esref{eqn: DIP volume element} to \eqref{eqn: IJK m flux}) can be evaluated analytically. For example, the flux elements $\mathcal{B}^+_{\alpha\beta\gamma\delta ijk}$ (\eref{eqn: Bp flux}) through boundaries of constant $r$ are given by \eref{eqn: Bp flux Schwarzchild}. 

\eref{eqn: Bp flux Schwarzchild} is not a particularly elegant expression but importantly it, and all other flux expressions, are independent of the distribution function and any state variables of the system. Therefore all terms relating to transport, fluxes and collisions, are pre-computable as fixed arrays.

\begin{widetext}
\begin{equation}
\begin{split}\label{eqn: Bp flux Schwarzchild}
    &\mathcal{B}^+_{\alpha\beta\gamma\delta ijk}=-\frac{1}{2}\left(t_{\alpha+1}-t_{\alpha}\right)\left(r^{3/2}_\beta\sqrt{r_\beta-r_s}\right)\left(\cos \theta_{\gamma+1}-\cos \theta_{\gamma}\right)\left(\psi_{\delta+1}-\psi_{\delta}\right)\left(\sqrt{m^2+p^2_{i+1}}-\sqrt{m^2+p^2_{i}}\right)
    \\
    &\times\left(u_{j+1}\sqrt{1-u_{j+1}^2}-2\text{arccot}\left(\frac{u_{j+1}-1}{\sqrt{1-u_{j+1}^2}}\right)-u_{j}\sqrt{1-u_{j}^2}+2\text{arccot}\left(\frac{u_{j}-1}{\sqrt{1-u_{j}^2}}\right)\right)\left(\sin\phi_{k+1}-\sin\phi_{k}\right).
\end{split}
\end{equation}
\end{widetext}

%% file: Sections/conclusion.tex
\section{Conclusion}\label{sec: conclusion}

This is the first paper in a series describing a novel mesoscopic framework for the general transport of particle distributions, with the aim of application to astrophysical systems.

In this paper, the transport of the particle distribution through phase space has been described in a general manner. Continuous forces and discrete particle interactions have been included (\ssref{sec:general} and \ref{sec:coordinate}).

This general description is independent of the underlying spacetime and agnostic to the types of particle interactions involved.

A novel approach of phase space discretisation of the particle distribution function, \textit{Distribution-In-Plateaux} (\sref{sec:DIP}), has been described that allows for the numerical evaluation of this transport in a conservative form. 

Numerical implementation and microscopic testing of this approach is detailed in \citetalias{PaperII}, with macroscopic tests to appear in \citetalias{PaperIII}. 

It is planned to apply the DIPLODOCUS framework to AGN jet modelling in the future. In particular, the added dimensionality allowed by the relaxations of various assumptions on the particle distributions and jet geometries will be exploited. This will allow their effects on emission spectra to be examined and may provide insight into the jet content and particle acceleration mechanisms.

%% file: Appendicies/absorptionanddecay.tex
\section{Collision Integral for Binary and Emissive interactions}\label{app: abs and emi}

\subsection{Binary Interactions}

Consider the reversible binary (scattering) interaction $\mathfrak{a}\mathfrak{b}\rightleftharpoons \mathfrak{c}\mathfrak{d}$. The collision integral for a particle of type-$\mathfrak{c}$ in this interaction is given by \eref{eqn: diff collision integral}: 
\begin{equation}
    \bm{C}(\bm{x},\bm{p}_\mathfrak{c})=\bm{N}_{\mathfrak{a}\mathfrak{b}\rightarrow\mathfrak{c}\mathfrak{d}}-\bm{N}_{\mathfrak{a}\mathfrak{b}\leftarrow \mathfrak{c}\mathfrak{d}}.
\end{equation} 
$\bm{N}_{\mathfrak{a}\mathfrak{b}\rightarrow\mathfrak{c}\mathfrak{d}}$ is the number of worldlines created by the forward interaction $\mathfrak{a}\mathfrak{b}\rightarrow\mathfrak{c}\mathfrak{d}$ in this reversible process, hence giving the number of particle type-$\mathfrak{c}$ (and type-$\mathfrak{d}$) worldlines started by that interaction within a volume of phase space. Using \eref{eqn: N worldlines} this may be written 
\begin{equation}
    \bm{N}_{\mathfrak{a}\mathfrak{b}\rightarrow\mathfrak{c}\mathfrak{d}}=\delta^{(4)}(\bm{p}_\mathfrak{a}+\bm{p}_\mathfrak{b}-\bm{p}_\mathfrak{c}-\bm{p}_\mathfrak{d})\left|M_{\mathfrak{a}\mathfrak{b}\rightarrow\mathfrak{c}\mathfrak{d}}\right|^2f_{\mathfrak{a}}(\bm{x},\bm{p}_\mathfrak{a})f_{\mathfrak{b}}(\bm{x},\bm{p}_\mathfrak{b})\frac{\bm{\pi}_{\mathfrak{a}}\bm{\pi}_{\mathfrak{b}}}{(1+\delta_{\mathfrak{a}\mathfrak{b}})}\frac{\bm{\pi}_{\mathfrak{c}}\bm{\pi}_{\mathfrak{d}}}{(1+\delta_{\mathfrak{c}\mathfrak{d}})}\bm{\chi}.
\end{equation}
Three of the Dirac delta functions in above equation can be removed by integrating over the momentum states of the other output particle in the interaction, i.e. states of type-$\mathfrak{d}$ particles 
\begin{equation}\label{eqn: binary N example}
    \bm{N}_{\mathfrak{a}\mathfrak{b}\rightarrow\mathfrak{c}\mathfrak{d}}=\delta(p^0_\mathfrak{a}+p^0_\mathfrak{b}-p^0_\mathfrak{c}-p^0_\mathfrak{d})\left|M_{\mathfrak{a}\mathfrak{b}\rightarrow\mathfrak{c}\mathfrak{d}}\right|^2f_{\mathfrak{a}}(\bm{x},\bm{p}_\mathfrak{a})f_{\mathfrak{b}}(\bm{x},\bm{p}_\mathfrak{b})\frac{\mathrm{d}^3p_\mathfrak{a}\mathrm{d}^3p_\mathfrak{b}}{p^0_\mathfrak{a}p^0_\mathfrak{b}(1+\delta_{\mathfrak{a}\mathfrak{b}})}\frac{\mathrm{d}^3p_\mathfrak{c}}{p^0_\mathfrak{c}p^0_\mathfrak{d}(1+\delta_{\mathfrak{c}\mathfrak{d}})}\bm{\chi},
\end{equation}
where $p^0_\mathfrak{d}$ is now a function of the other particle momenta given by energy conservation $p^0_\mathfrak{d}=p^0_\mathfrak{a}+p^0_\mathfrak{b}-p^0_\mathfrak{c}$ and note the shorthand for the momentum space volume element $\mathrm{d}^3p/p^0=(p)^2\mathrm{d}p\mathrm{d}u\mathrm{d}\phi/p^0=\bm{\mathrm{d}}p^1\wedge\bm{\mathrm{d}}p^2\wedge\bm{\mathrm{d}}p^3/p^0$.

For binary collisions, the rate is typically measured using a differential cross section. For generality of reference frame, it is best to use the Lorentz invariant differential cross section $\frac{\mathrm{d}\sigma_{\mathfrak{a}\mathfrak{b}\rightarrow \mathfrak{c}\mathfrak{d}}}{\mathrm{d}T}$, which is a function of the standard Mandelstam variables $S,T\text{ and }U$. This differential cross section can be related to the scattering matrix (\eref{eqn: T to M scattering}) by
\begin{equation}
    \left|M_{\mathfrak{a}\mathfrak{b}\rightarrow\mathfrak{c}\mathfrak{d}}\right|^2=\frac{\mathcal{F}_{\mathfrak{a}\mathfrak{b}}^2}{\pi}\frac{\mathrm{d}\sigma_{\mathfrak{a}\mathfrak{b}\rightarrow\mathfrak{c}\mathfrak{d}}}{\mathrm{d}T},
\end{equation}
where $\mathcal{F}_{\mathfrak{a}\mathfrak{b}}=\frac{1}{2}\sqrt{\left(S-(m_\mathfrak{a}+m_\mathfrak{b})^2\right)\left(S-(m_\mathfrak{a}-m_\mathfrak{b})^2\right)}$ is a Lorentz scalar related to the incident flux of particles of type-$\mathfrak{a}$ and type-$\mathfrak{b}$.

The internal terms of \eref{eqn: binary N example} may then be collected into the expression 
\begin{equation}\label{eqn: binary gain term}
    \bm{N}_{\mathfrak{a}\mathfrak{b}\rightarrow\mathfrak{c}\mathfrak{d}}=f_{\mathfrak{a}}(\bm{x},\bm{p}_\mathfrak{a})f_{\mathfrak{b}}(\bm{x},\bm{p}_\mathfrak{b})G_{\mathfrak{a}\mathfrak{b}\rightarrow\mathfrak{c}\mathfrak{d}}\mathrm{d}^3p_\mathfrak{a}\mathrm{d}^3p_\mathfrak{b}\mathrm{d}^3p_\mathfrak{c}\bm{\chi}, 
    \quad
    G_{\mathfrak{a}\mathfrak{b}\rightarrow\mathfrak{c}\mathfrak{d}}=\frac{\delta(p^0_\mathfrak{a}+p^0_\mathfrak{b}-p^0_\mathfrak{c}-p^0_\mathfrak{d})}{(1+\delta_{\mathfrak{a}\mathfrak{b}})(1+\delta_{\mathfrak{c}\mathfrak{d}})}\frac{\mathcal{F}_{\mathfrak{a}\mathfrak{b}}^2}{\pi p^0_\mathfrak{a}p^0_\mathfrak{b}p^0_\mathfrak{c}p^0_\mathfrak{d}}\frac{\mathrm{d}\sigma_{\mathfrak{a}\mathfrak{b}\rightarrow\mathfrak{c}\mathfrak{d}}}{\mathrm{d}T},
\end{equation}
with $G_{\mathfrak{a}\mathfrak{b}\rightarrow\mathfrak{c}\mathfrak{d}}$ being denoted the ``gain term'' of particles of type-$\mathfrak{c}$ (and type-$\mathfrak{d}$) for the forward reaction.

For the reverse process $\bm{N}_{\mathfrak{a}\mathfrak{b}\leftarrow \mathfrak{c}\mathfrak{d}}$, \eref{eqn: binary gain matrix} can be altered and simplified by acknowledging that now both output states (now particles of type-$\mathfrak{a}$ and type-$\mathfrak{b}$) can be integrated over, leading to 
\begin{equation} \label{eqn: binary loss term}
    \bm{N}_{\mathfrak{a}\mathfrak{b}\leftarrow\mathfrak{c}\mathfrak{d}}=f_{\mathfrak{c}}(\bm{x},\bm{p}_\mathfrak{c})f_{\mathfrak{d}}(\bm{x},\bm{p}_\mathfrak{d})L_{\mathfrak{a}\mathfrak{b}\leftarrow\mathfrak{c}\mathfrak{d}}\mathrm{d}^3p_\mathfrak{c}\mathrm{d}^3p_\mathfrak{d}\bm{\chi},
    \quad
    L_{\mathfrak{a}\mathfrak{b}\leftarrow\mathfrak{c}\mathfrak{d}} = \frac{1}{\left(1+\delta_{\mathfrak{c}\mathfrak{d}}\right)}\frac{\mathcal{F}_{\mathfrak{c}\mathfrak{d}}\sigma_{\mathfrak{a}\mathfrak{b}\leftarrow\mathfrak{c}\mathfrak{d}}}{p^0_\mathfrak{c}p^0_\mathfrak{d}},
\end{equation}
where $\sigma_{\mathfrak{a}\mathfrak{b}\leftarrow\mathfrak{c}\mathfrak{d}}$ is the Lorentz invariant total cross section for the reverse process $\mathfrak{a}\mathfrak{b}\leftarrow \mathfrak{c}\mathfrak{d}$, which typically involves the factor of $1/(1+\delta_{\mathfrak{a}\mathfrak{b}})$ in its evaluation, and $L_{\mathfrak{a}\mathfrak{b}\leftarrow\mathfrak{c}\mathfrak{d}}$ is the ``loss term'' for particles of type-$\mathfrak{c}$ (and type-$\mathfrak{d}$) for the reverse reaction.

\subsection{Binary Interactions in DIP Form}
Under the assumption that the distribution function takes the form granted by DIP, \eref{eqn: DIP}, the gain and loss terms \esref{eqn: binary gain term} and \eqref{eqn: binary loss term} may be integrated over sub-domains of the momentum space of the particles involved to give the ``gain array'' components:
\begin{equation}\label{eqn: binary gain matrix}
    G_{\mathfrak{a}\mathfrak{b}\rightarrow\mathfrak{c}\mathfrak{d},ijklmnopq}=\int_{P_{ijk}}\int_{P_{lmn}}\int_{P_{opq}}G_{\mathfrak{a}\mathfrak{b}\rightarrow\mathfrak{c}\mathfrak{d}}\frac{\mathrm{d}^3p_\mathfrak{b}}{(p_\mathfrak{b})^2\Delta p_{\mathfrak{b},o}\Delta u_{\mathfrak{b},p}\Delta \phi_{\mathfrak{b},q}}\frac{\mathrm{d}^3p_\mathfrak{a}}{(p_\mathfrak{a})^2\Delta p_{\mathfrak{a},l}\Delta u_{\mathfrak{a},m}\Delta \phi_{\mathfrak{a},n}}\mathrm{d}^3p_\mathfrak{c}
\end{equation}
and ``loss array'' components
\begin{equation}\label{eqn: binary loss matrix}
    L_{\mathfrak{a}\mathfrak{b}\leftarrow\mathfrak{c}\mathfrak{d},ijklmn} = \int_{P_{ijk}}\int_{P_{lmn}}L_{\mathfrak{a}\mathfrak{b}\leftarrow\mathfrak{c}\mathfrak{d}}\frac{\mathrm{d}^3p_\mathfrak{d}}{(p_\mathfrak{d})^2\Delta p_{\mathfrak{d},l}\Delta u_{\mathfrak{d},m}\Delta \phi_{\mathfrak{d},n}}\frac{\mathrm{d}^3p_\mathfrak{c}}{(p_\mathfrak{c})^2\Delta p_{\mathfrak{c},i}\Delta u_{\mathfrak{c},j}\Delta \phi_{\mathfrak{c},k}}
\end{equation}
which are pre-computable by numerical methods (see \citetalias{PaperII}) and independent of spatial coordinates, and as such are valid at any point in space.

\subsection{Emissive Interactions}
Consider an emissive interaction where one particle ``decays'' into two $\mathfrak{a}\rightarrow\mathfrak{b}\mathfrak{c}$. The collision integral for a particle of type-$\mathfrak{c}$ in this interaction is given by \eref{eqn: diff collision integral}:
\begin{equation}
    \bm{C}(\bm{x},\bm{p}_\mathfrak{c})=\bm{N}_{\mathfrak{a}\rightarrow\mathfrak{b}\mathfrak{c}},
\end{equation}
$\bm{N}_{\mathfrak{a}\rightarrow\mathfrak{b}\mathfrak{c}}$ is the number of worldline of particles of type-$\mathfrak{c}$ created by the emissive interaction for $\mathfrak{a}\rightarrow\mathfrak{b}\mathfrak{c}$. Using \eref{eqn: N worldlines} this may be written:
\begin{equation}
    \bm{N}_{\mathfrak{a}\rightarrow\mathfrak{b}\mathfrak{c}} = \delta^{(4)}(\bm{p}_\mathfrak{a} - \bm{p}_\mathfrak{b} - \bm{p}_\mathfrak{c}) \left| M_{\mathfrak{a}\rightarrow\mathfrak{b}\mathfrak{c}} \right|^2 f_{\mathfrak{a}}(\bm{x}, \bm{p}_\mathfrak{a}) \bm{\pi}_{\mathfrak{a}} \frac{\bm{\pi}_{\mathfrak{b}} \bm{\pi}_{\mathfrak{c}}}{(1+\delta_{\mathfrak{b}\mathfrak{c}})} \bm{\chi}.
\end{equation}
Three of the Dirac delta functions in the above equation can be removed by integrating over the momentum states of the outgoing particle of type-$\mathfrak{b}$:
\begin{equation}\label{eqn: emission gain term}
    \bm{N}_{\mathfrak{a}\rightarrow\mathfrak{b}\mathfrak{c}} = \frac{\delta(p^0_\mathfrak{a} - p^0_\mathfrak{b} - p^0_\mathfrak{c}) \left| M_{\mathfrak{a}\rightarrow\mathfrak{b}\mathfrak{c}} \right|^2}{p^0_\mathfrak{a}p^0_\mathfrak{b}p^0_\mathfrak{c}(1+\delta_{\mathfrak{b}\mathfrak{c}})} f_{\mathfrak{a}}(\bm{x}, \bm{p}_\mathfrak{a})  \mathrm{d}^3p_\mathfrak{a}\mathrm{d}^3p_\mathfrak{c} \bm{\chi}=\frac{\mathrm{d}N_{\mathfrak{a}\rightarrow\mathfrak{b}\mathfrak{c}}}{\mathrm{d}x^0\mathrm{d}^3p_\mathfrak{c}}f_{\mathfrak{a}}(\bm{x}, \bm{p}_\mathfrak{a}) \mathrm{d}^3p_\mathfrak{a}\mathrm{d}^3p_\mathfrak{c} \bm{\chi},
\end{equation}
where $\frac{\mathrm{d}N_{\mathfrak{a}\rightarrow\mathfrak{b}\mathfrak{c}}}{\mathrm{d}x^0\mathrm{d}^3p_\mathfrak{c}}$ is the single particle emissivity of the interaction. This single particle emissivity is typically dependent on some external field such as the presence of a magnetic field in the case of synchrotron emissions.

\subsection{Emissive Interactions in DIP Form}
Under the assumptions that the distribution of the particles takes the form granted by DIP, \eref{eqn: DIP}, the gain array for an emissive interaction $\mathfrak{a}\rightarrow\mathfrak{b}\mathfrak{c}$ can be generated by integrating \eref{eqn: emission gain term} over discrete sub-domains of momentum space:
\begin{equation}
    G_{\mathfrak{a}\rightarrow\mathfrak{b}\mathfrak{c},ijklmn}=\int_{P_{ijk}}\int_{P_{lmn}}G_{\mathfrak{a}\rightarrow\mathfrak{b}\mathfrak{c}}\frac{\mathrm{d}^3p_\mathfrak{a}}{(p_\mathfrak{a})^2\Delta p_{\mathfrak{a},l}\Delta u_{\mathfrak{a},m}\Delta \phi_{\mathfrak{a},n}}\frac{\mathrm{d}^3p_\mathfrak{c}}{(p_\mathfrak{c})^2}, \quad G_{\mathfrak{a}\rightarrow\mathfrak{b}\mathfrak{c}} = (p_\mathfrak{c})^2\frac{\mathrm{d}N_{\mathfrak{a}\rightarrow\mathfrak{b}\mathfrak{c}}}{\mathrm{d}x^0\mathrm{d}^3p_\mathfrak{c}} 
\end{equation}

%% file: Appendicies/metrics.tex
\section{Metrics}\label{app:metrics}

A metric of the form \eref{eqn: stationary metric general}, can express most standard astrophycially relevant metrics. A subset are presented below:

Cartesian Minkowski:
\begin{equation}
      x^\alpha=\{t,x,y,z\},~ A^2=1,~B^2=1,~C^2=1,~D^2=1,~E=F=G=0.
\end{equation}

Spherical Minkowski:
\begin{equation} 
    x^\alpha=\{t,r,\theta,\psi\},~ A^2=1,~B^2=1,~C^2=r^2,~D^2=r^2\sin^2\theta,~E=F=G=0.
\end{equation}

Cylindrical Minkowski:
\begin{equation}
    x^\alpha=\{t,\rho,\vartheta,z\},~ A^2=1,~B^2=1,~C^2=\rho^2,~D^2=1,~E=F=G=0.
\end{equation}

Schwarzschild:
\begin{equation}\label{eqn: schw coordinates}
        \quad  x^\alpha=\{t,r,\theta,\psi\},~ A^2=\left(1-\frac{r_s}{r}\right),~B^2=\left(1-\frac{r_s}{r}\right)^{-1},~C^2=r^2,~D^2=r^2\sin^2\theta,~E=F=G=0.
\end{equation}

Kerr (Boyer-Lindquist Coordinates):
\begin{equation}
\begin{split} 
    x^\alpha &= \{t,r,\theta,\psi\}, ~A^2=\left(1-\frac{r_s}{r}\right),~B^2=\frac{\Sigma}{\Delta},~C^2=\Sigma,~D^2=\left(r^2+a^2+\frac{r_sra^2}{\Sigma}\sin^2\theta\right)\sin^2\theta,~E=F=0,
    \\
    &\quad\quad  G=\frac{-r_s r a\sin^2\theta}{\Sigma},~\Sigma=r^2+a^2\cos^2\theta,~\Delta = r^2-rr_s+a^2.
\end{split}
\end{equation}

%% file: Appendicies/longequations.tex
\section{DIP Transport Equations}\label{app:longeq}

Under the assumption that the distribution function $f(\bm{x},\bm{p})$ for all particles takes the DIP form, \eref{eqn: DIP}, the integral form of the transport equation \eref{eqn: integral transport w coll}, for a particle of type-$\mathfrak{c}$ can be evaluated over a phase-space sub-domain $X_{\alpha\beta\gamma\delta}\times P_{ijk}=[t_\alpha,t_{\alpha+1}]\times[x_\beta,x_{\beta+1}]\times[y_\gamma,y_{\gamma+1}]\times[z_\delta,z_{\delta+1}]\times[p_{i},p_{i+1}]\times[u_{j},u_{j+1}]\times[\phi_{k},\phi_{k+1}]$ to give the discrete DIP transport equation:
\begin{equation}\label{eqn: full transport DIP}
    \begin{split}
    & \left(h^-_{\alpha+1}f_{\mathfrak{c},(\alpha+1)\beta\gamma\delta ijk}+h^+_{\alpha}f_{\mathfrak{c},\alpha\beta\gamma\delta ijk}\right)\mathcal{A}^{+}_{\alpha\beta\gamma\delta ijk}
    +
    \left(h^-_{\alpha}f_{\mathfrak{c},\alpha\beta\gamma\delta ijk}+h^+_{\alpha-1}f_{\mathfrak{c},(\alpha-1)\beta\gamma\delta ijk}\right)\mathcal{A}^{-}_{\alpha\beta\gamma\delta ijk}
    \\ &\quad\quad
    +
    \left(h^-_{\beta+1}f_{\mathfrak{c},\alpha(\beta+1)\gamma\delta ijk}-h^+_{\beta}f_{\mathfrak{c},\alpha\beta\gamma\delta ijk}\right)\mathcal{B}^{+}_{\alpha\beta\gamma\delta ijk}
    +
    \left(h^-_{\beta}f_{\mathfrak{c},\alpha\beta\gamma\delta ijk}-h^+_{\beta-1}f_{\mathfrak{c},\alpha(\beta-1)\gamma\delta ijk}\right)\mathcal{B}^{-}_{\alpha\beta\gamma\delta ijk}
    \\ &\quad\quad
    +
    \left(h^-_{\gamma+1}f_{\mathfrak{c},\alpha\beta(\gamma+1)\delta ijk}+h^+_{\gamma}f_{\mathfrak{c},\alpha\beta\gamma\delta ijk}\right)\mathcal{C}^{+}_{\alpha\beta\gamma\delta ijk}
    +
    \left(h^-_{\gamma}f_{\mathfrak{c},\alpha\beta\gamma\delta ijk}+h^+_{\gamma-1}f_{\mathfrak{c},\alpha\beta(\gamma-1)\delta ijk}\right)\mathcal{C}^{-}_{\alpha\beta\gamma\delta ijk}
    \\ &\quad\quad
    +
    \left(h^-_{\delta+1}f_{\mathfrak{c},\alpha\beta\gamma(\delta+1) ijk}+h^+_{\delta}f_{\mathfrak{c},\alpha\beta\gamma\delta ijk}\right)\mathcal{D}^{+}_{\alpha\beta\gamma\delta ijk}
    +
    \left(h^-_{\delta}f_{\mathfrak{c},\alpha\beta\gamma\delta ijk}+h^+_{\delta-1}f_{\mathfrak{c},\alpha\beta\gamma(\delta-1) ijk}\right)\mathcal{D}^{-}_{\alpha\beta\gamma\delta ijk}
    \\ &\quad\quad
    +
    \left(\frac{h^-_{i+1}f_{\mathfrak{c},\alpha\beta\gamma\delta(i+1)jk}}{\Delta p_{\mathfrak{c},i+1}}+\frac{h^+_{i}f_{\mathfrak{c},\alpha\beta\gamma\delta ijk}}{\Delta p_{\mathfrak{c},i}}\right)\mathcal{I}^{+}_{\alpha\beta\gamma\delta ijk}
    +
    \left(\frac{h^-_{i}f_{\mathfrak{c},\alpha\beta\gamma\delta ijk}}{\Delta p_{\mathfrak{c},i}}+\frac{h^+_{i-1}f_{\mathfrak{c},\alpha\beta\gamma\delta(i-1)jk}}{\Delta p_{\mathfrak{c},i-1}}\right)\mathcal{I}^{-}_{\alpha\beta\gamma\delta ijk}
    \\ &\quad\quad
    +
    \left(\frac{h^-_{j+1}f_{\mathfrak{c},\alpha\beta\gamma\delta i(j+1)k}}{\Delta u_{\mathfrak{c},j+1}}+\frac{h^+_{j}f_{\mathfrak{c},\alpha\beta\gamma\delta ijk}}{\Delta u_{\mathfrak{c},j}}\right)\mathcal{J}^{+}_{\alpha\beta\gamma\delta ijk}
    +
    \left(\frac{h^-_{j}f_{\mathfrak{c},\alpha\beta\gamma\delta ijk}}{\Delta u_{\mathfrak{c},j}}+\frac{h^+_{j-1}f_{\mathfrak{c},\alpha\beta\gamma\delta i(j-1)k}}{\Delta u_{\mathfrak{c},j-1}}\right)\mathcal{J}^{-}_{\alpha\beta\gamma\delta ijk}
    \\ &\quad\quad
    +
    \left(\frac{h^-_{k+1}f_{\mathfrak{c},\alpha\beta\gamma\delta ij(k+1)}}{\Delta \phi_{\mathfrak{c},k+1}}+\frac{h^+_{k}f_{\mathfrak{c},\alpha\beta\gamma\delta ijk}}{\Delta \phi_{\mathfrak{c},k}}\right)\mathcal{K}^{+}_{\alpha\beta\gamma\delta ijk}
    +
    \left(\frac{h^-_{k}f_{\mathfrak{c},\alpha\beta\gamma\delta ijk}}{\Delta \phi_{\mathfrak{c},k}}+\frac{h^+_{k-1}f_{\mathfrak{c},\alpha\beta\gamma\delta ij(k-1)}}{\Delta \phi_{\mathfrak{c},k-1}}\right)\mathcal{K}^{-}_{\alpha\beta\gamma\delta ijk}
    \\
    &=\sum_{\mathfrak{a}\mathfrak{b}\mathfrak{d}\in \text{particles}}\left(\sum_{lmnopq}G_{\mathfrak{a}\mathfrak{b}\rightarrow\mathfrak{c}\mathfrak{d},ijklmnopq}f_{\mathfrak{a},\alpha\beta\gamma\delta lmn}f_{\mathfrak{b},\alpha\beta\gamma\delta opq}-\sum_{lmn}L_{\mathfrak{a}\mathfrak{b}\leftarrow\mathfrak{c}\mathfrak{d},ijklmn}f_{\mathfrak{c},\alpha\beta\gamma\delta ijk}f_{\mathfrak{d},\alpha\beta\gamma\delta lmn}\right)\mathcal{V}_{\alpha\beta\gamma\delta}
    \\
    &\quad\quad 
    +
    \sum_{\mathfrak{a}\mathfrak{b}\in \text{particles}}\sum_{lmn}G_{\mathfrak{a}\rightarrow\mathfrak{b}\mathfrak{c},ijklmn}f_{\mathfrak{a},\alpha\beta\gamma\delta lmn} \mathcal{V}_{\alpha\beta\gamma\delta}
    \end{split}
\end{equation}
where the spacetime volume element is
\begin{equation}\label{eqn: DIP volume element}
    \mathcal{V}_{\alpha\beta\gamma\delta}=\int_{z_{\delta}}^{z_{\delta+1}}\int_{y_{\gamma}}^{y_{\gamma+1}}\int_{x_{\beta}}^{x_{\beta+1}}\int_{t_{\alpha}}^{t_{\alpha+1}} \chi_{0123}\mathrm{d}t\mathrm{d}x\mathrm{d}y\mathrm{d}z,
\end{equation}
the fluxes through hypersurfaces of constant time coordinate are:
\begin{align}
    \label{eqn: Ap flux}
    \mathcal{A}^{+}_{\alpha\beta\gamma\delta ijk}&=\int_{\phi_{k}}^{\phi_{k+1}}\int_{u_{j}}^{u_{j+1}}\int_{p_{i}}^{p_{i+1}}\int_{z_{\delta}}^{z_{\delta+1}}\int_{y_{\gamma}}^{y_{\gamma+1}}\int_{x_{\beta}}^{x_{\beta+1}} \left[p^ae_a^{~0}\frac{\chi_{0123}}{p^0}\right]_{t_{\alpha+1}}\frac{\mathrm{d}x\mathrm{d}y\mathrm{d}z\mathrm{d}p\mathrm{d}u\mathrm{d}\phi}{\Delta p_i\Delta u_j\Delta \phi_k}, 
    \\
    \label{eqn: Am flux}
    \mathcal{A}^{-}_{\alpha\beta\gamma\delta ijk} &= -\mathcal{A}^{+}_{(\alpha-1)\beta\gamma\delta ijk},
\end{align}
the fluxes through hypersurfaces of constant spatial coordinates are:
\begin{align}
    \label{eqn: Bp flux}
    \mathcal{B}^{+}_{\alpha\beta\gamma\delta ijk}&=\int_{\phi_{k}}^{\phi_{k+1}}\int_{u_{j}}^{u_{j+1}}\int_{p_{i}}^{p_{i+1}}\int_{t_{\alpha}}^{t_{\alpha+1}}\int_{z_{\delta}}^{z_{\delta+1}}\int_{y_{\gamma}}^{y_{\gamma+1}}  \left[p^ae_a^{~1}\frac{\chi_{1230}}{p^0}\right]_{x_{\beta+1}} \frac{\mathrm{d}y\mathrm{d}z\mathrm{d}t\mathrm{d}p\mathrm{d}u\mathrm{d}\phi}{\Delta p_i\Delta u_j\Delta \phi_k}, \quad
    \\
    \label{eqn: Cp flux}
    \mathcal{C}^{+}_{\alpha\beta\gamma\delta ijk}&=\int_{\phi_{k}}^{\phi_{k+1}}\int_{u_{j}}^{u_{j+1}}\int_{p_{i}}^{p_{i+1}}\int_{x_{\beta}}^{x_{\beta+1}}\int_{t_{\alpha}}^{t_{\alpha+1}}\int_{z_{\delta}}^{z_{\delta+1}}\left[p^ae_a^{~2}\frac{\chi_{2301}}{p^0}\right]_{y_{\gamma+1}} \frac{\mathrm{d}z\mathrm{d}t\mathrm{d}x\mathrm{d}p\mathrm{d}u\mathrm{d}\phi}{\Delta p_i\Delta u_j\Delta \phi_k}, \quad 
    \\
    \label{eqn: Dp flux}
    \mathcal{D}^{+}_{\alpha\beta\gamma\delta ijk}&=\int_{\phi_{k}}^{\phi_{k+1}}\int_{u_{j}}^{u_{j+1}}\int_{p_{i}}^{p_{i+1}}\int_{y_{\gamma}}^{y_{\gamma+1}}\int_{x_{\beta}}^{x_{\beta+1}}\int_{t_{\alpha}}^{t_{\alpha+1}}\left[p^ae_a^{~3}\frac{\chi_{3012}}{p^0}\right]_{z_{\delta+1}} \frac{\mathrm{d}t\mathrm{d}x\mathrm{d}y\mathrm{d}p\mathrm{d}u\mathrm{d}\phi}{\Delta p_i\Delta u_j\Delta \phi_k}, \quad 
    \\
    \label{eqn: BCD m flux}
    \mathcal{B}^{-}_{\alpha\beta\gamma\delta ijk} &= -\mathcal{B}^{+}_{\alpha(\beta-1)\gamma\delta ijk}, 
    \quad
    \mathcal{C}^{-}_{\alpha\beta\gamma\delta ijk} = -\mathcal{C}^{+}_{\alpha\beta(\gamma-1)\delta ijk},
    \quad
    \mathcal{D}^{-}_{\alpha\beta\gamma\delta ijk}=-\mathcal{D}^{+}_{\alpha\beta\gamma(\delta-1) ijk}, 
\end{align}
the fluxes through constant momentum coordinates are:
\begin{align}
    \label{eqn: Ip flux}
    \mathcal{I}^{+}_{\alpha\beta\gamma\delta ijk}&=\int_{\phi_{k}}^{\phi_{k+1}}\int_{u_{j}}^{u_{j+1}}\int_{z_{\delta}}^{z_{\delta+1}}\int_{y_{\gamma}}^{y_{\gamma+1}}\int_{x_{\beta}}^{x_{\beta+1}}\int_{t_{\alpha}}^{t_{\alpha+1}} \left[\frac{\chi_{0123}}{p^0}R^1_{~a}\left(-\Gamma^{a}_{~bc}p^bp^c+mF^a\right)\right]_{p_{i+1}} \frac{\mathrm{d}t\mathrm{d}x\mathrm{d}y\mathrm{d}z\mathrm{d}u\mathrm{d}\phi}{\Delta u_j\Delta \phi_k},
    \\
    \label{eqn: Jp flux}
    \mathcal{J}^{+}_{\alpha\beta\gamma\delta ijk}&=\int_{p_{i}}^{p_{i+1}}\int_{\phi_{k}}^{\phi_{k+1}}\int_{z_{\delta}}^{z_{\delta+1}}\int_{y_{\gamma}}^{y_{\gamma+1}}\int_{x_{\beta}}^{x_{\beta+1}}\int_{t_{\alpha}}^{t_{\alpha+1}} \left[\frac{\chi_{0123}}{p^0}R^2_{~a}\left(-\Gamma^{a}_{~bc}p^bp^c+mF^a\right)\right]_{u_{j+1}} \frac{\mathrm{d}t\mathrm{d}x\mathrm{d}y\mathrm{d}z\mathrm{d}\phi\mathrm{d}p}{\Delta p_i\Delta \phi_k},
    \\
    \label{eqn: Kp flux}
    \mathcal{K}^{+}_{\alpha\beta\gamma\delta ijk}&=\int_{u_{j}}^{u_{j+1}}\int_{p_{i}}^{p_{i+1}}\int_{z_{\delta}}^{z_{\delta+1}}\int_{y_{\gamma}}^{y_{\gamma+1}}\int_{x_{\beta}}^{x_{\beta+1}}\int_{t_{\alpha}}^{t_{\alpha+1}} \left[\frac{\chi_{0123}}{p^0}R^3_{~a}\left(-\Gamma^{a}_{~bc}p^bp^c+mF^a\right)\right]_{\phi_{k+1}} \frac{\mathrm{d}t\mathrm{d}x\mathrm{d}y\mathrm{d}z\mathrm{d}p\mathrm{d}u}{\Delta p_i\Delta u_j},
    \\
    \label{eqn: IJK m flux}
    \mathcal{I}^{-}_{\alpha\beta\gamma\delta ijk} &= -\mathcal{I}^{+}_{\alpha\beta\gamma\delta (i-1)jk},
    \quad 
    \mathcal{J}^{-}_{\alpha\beta\gamma\delta ijk} = -\mathcal{J}^{+}_{\alpha\beta\gamma\delta i(j-1)k},
    \quad 
    \mathcal{K}^{-}_{\alpha\beta\gamma\delta ijk} = -\mathcal{K}^{+}_{\alpha\beta\gamma\delta ij(k-1)},
\end{align}
with $p^a=(p^0,p\sqrt{1-u^2}\cos\phi,p\sqrt{1-u^2}\sin\phi,pu)$ in modified spherical coordinates, with the associated transformation for vector components 
\begin{equation}\label{eqn: rotation transformation matrix}
    R^a_{~b}=\begin{pmatrix}
        1 & 0 & 0 & 0 \\
        0 & \sqrt{1-u^2}\cos\phi & \sqrt{1-u^2}\sin\phi & u \\
        0 & -\frac{u\sqrt{1-u^2}\cos\phi}{p}  & -\frac{u\sqrt{1-u^2}\sin\phi}{p}  &  \frac{1-u^2}{p}  \\
        0 & -\frac{\sin\phi}{p\sqrt{1-u^2}}  & \frac{\cos\phi}{p\sqrt{1-u^2}}  &  0   \\
    \end{pmatrix}.
\end{equation} 

The volume elements and fluxes (\esref{eqn: DIP volume element} to \eqref{eqn: IJK m flux}) are all integrals over phase-space sub-domains, which in most cases can be analytically integrated as the form of all the internal functions, connection coefficients, external forces etc. are known (see, for example, \eref{eqn: Bp flux Schwarzchild}).

%% file: references.bib
@book{BerestetskiiEtAl_1982,
  title = {Quantum Electrodynamics},
  author = {Berestetskii, V. B. and Lifshits, E. M. and Pitaevskii, L. P.},
  translator = {Sykes, J. B. and Bell, S. J.},
  year = {1982},
  edition = {2nd ed.},
  publisher = {{Butterworth-Heinemann}},
  isbn = {978-0-08-050346-2},
  langid = {english}
}

@article{StepneyGuilbert_1983,
    author = {{Stepney}, S. and {Guilbert}, P.~W.},
    title = "{Numerical fits to important rates in high temperature astrophysical plasmas.}",
    journal = {\mnras},
    year = 1983,
    month = sep,
    volume = {204},
    pages = {1269-1277},
    doi = {10.1093/mnras/204.4.1269},
    adsurl = {https://ui.adsabs.harvard.edu/abs/1983MNRAS.204.1269S},
}

@article{PotterCotter_2012,
  title = {Synchrotron and Inverse-{{Compton}} Emission from Blazar Jets - {{I}}. {{A}} Uniform Conical Jet Model: {{A}} Uniform Conical Jet Model},
  shorttitle = {Synchrotron and Inverse-{{Compton}} Emission from Blazar Jets - {{I}}. {{A}} Uniform Conical Jet Model},
  author = {Potter, William J. and Cotter, Garret},
  year = {2012},
  journal = {\mnras},
  volume = {423},
  number = {1},
  pages = {756--765},
  issn = {00358711},
  doi = {10.1111/j.1365-2966.2012.20918.x},
       adsurl = {https://ui.adsabs.harvard.edu/abs/2012MNRAS.423..756P},
}

@article{PotterCotter_2013a,
  title = {Synchrotron and Inverse-{{Compton}} Emission from Blazar Jets - {{II}}. {{An}} Accelerating Jet Model with a Geometry Set by Observations of {{M87}}},
  author = {Potter, William J. and Cotter, Garret},
  year = {2013},
  journal = {\mnras},
  volume = {429},
  pages = {1189--1205},
  issn = {0035-8711},
  doi = {10.1093/mnras/sts407},
       adsurl = {https://ui.adsabs.harvard.edu/abs/2013MNRAS.429.1189P},
}

@article{PotterCotter_2013b,
  title = {Synchrotron and Inverse-{{Compton}} Emission from Blazar Jets - {{III}}. {{Compton-dominant}} Blazars},
  author = {Potter, William J. and Cotter, Garret},
  year = {2013},
  journal = {\mnras},
  volume = {431},
  pages = {1840--1852},
  issn = {0035-8711},
  doi = {10.1093/mnras/stt300},
       adsurl = {https://ui.adsabs.harvard.edu/abs/2013MNRAS.431.1840P},
}

@article{PotterCotter_2013c,
  title = {Synchrotron and Inverse-{{Compton}} Emission from Blazar Jets - {{IV}}. {{BL Lac}} Type Blazars and the Physical Basis for the Blazar Sequence},
  author = {Potter, William J. and Cotter, Garret},
  year = {2013},
  journal = {\mnras},
  volume = {436},
  pages = {304--314},
  issn = {0035-8711},
  doi = {10.1093/mnras/stt1569},
       adsurl = {https://ui.adsabs.harvard.edu/abs/2013MNRAS.436..304P},
}

@article{PotterCotter_2015,
  title = {New Constraints on the Structure and Dynamics of Black Hole Jets},
  author = {Potter, William J. and Cotter, Garret},
  year = 2015,
  journal = {\mnras},
  volume = {453},
  number = {4},
  pages = {4070--4088},
  issn = {0035-8711},
  doi = {10.1093/mnras/stv1657},
       adsurl = {https://ui.adsabs.harvard.edu/abs/2015MNRAS.453.4070P},
}

@article{UrryPadovani_1995,
  title = {Unified {{Schemes}} for {{Radio-Loud Active Galactic Nuclei}}},
  author = {Urry, C. Megan and Padovani, Paolo},
  year = 1995,
  journal = {\pasp},
  volume = {107},
  pages = {803},
  issn = {0004-6280, 1538-3873},
  doi = {10.1086/133630},
       adsurl = {https://ui.adsabs.harvard.edu/abs/1995PASP..107..803U},
}

@article{Baring_1987,
  title = {Reaction Rates and Spectra in Relativistic Plasmas},
  author = {Baring, Matthew},
  year = 1987,
  date = {1987-10-01},
  journal = {\mnras},
  volume = {228},
  number = {3},
  pages = {681--693},
  issn = {0035-8711},
  doi = {10.1093/mnras/228.3.681},
  adsurl = {https://ui.adsabs.harvard.edu/abs/1987MNRAS.228..681B}
}

@article{Svensson_1982,
  title = {The Pair Annihilation Process in Relativistic Plasmas},
  author = {Svensson, R.},
  date = {1982-07-01},
  year = 1982,
  journal = {\apj},
  volume = {258},
  pages = {321--334},
  issn = {0004-637X},
  doi = {10.1086/160081},
       adsurl = {https://ui.adsabs.harvard.edu/abs/1982ApJ...258..321S},
}

@article{Dermer_1984,
  title = {The Production Spectrum of a Relativistic {{Maxwell-Boltzmann}} Gas},
  author = {Dermer, C. D.},
  date = {1984-05-01},
  year = 1984,
  journal = {\apj},
  volume = {280},
  pages = {328--333},
  issn = {0004-637X},
  doi = {10.1086/161999},
adsurl = {https://ui.adsabs.harvard.edu/abs/1984ApJ...280..328D},
}

@article{Dermer_1985,
  title = {Binary Collision Rates of Relativistic Thermal Plasmas. {{I Theoretical}} Framework},
  author = {Dermer, C. D.},
  date = {1985-08-01},
  year = 1985,
  journal = {\apj},
  volume = {295},
  pages = {28--37},
  issn = {0004-637X},
  doi = {10.1086/163345},
       adsurl = {https://ui.adsabs.harvard.edu/abs/1985ApJ...295...28D},
}

@article{Dermer_1986,
  title = {Binary {{Collision Rates}} of {{Relativistic Thermal Plasmas}}. {{II}}. {{Spectra}}},
  author = {Dermer, C. D.},
  date = {1986-08-01},
  year = 1986,
  journal = {\apj},
  volume = {307},
  pages = {47},
  issn = {0004-637X},
  doi = {10.1086/164391},
       adsurl = {https://ui.adsabs.harvard.edu/abs/1986ApJ...307...47D}
}

@article{CoppiBlandford_1990,
  title = {Reaction Rates and Energy Distributions for Elementary Processes in Relativistic Pair Plasmas},
  author = {Coppi, P. S. and Blandford, R. D.},
  date = {1990-08-01},
  year = 1990,
  journal = {\mnras},
  volume = {245},
  pages = {453--453},
  issn = {0035-8711},
  doi = {10.1093/mnras/245.3.453},
  adsurl = {https://ui.adsabs.harvard.edu/abs/1990MNRAS.245..453C},
}

@book{GrootEtAl_1980,
  title = {Relativistic Kinetic Theory: Principles and Applications},
  shorttitle = {Relativistic Kinetic Theory},
  author = {Groot, S. R. and Leeuwen, W. A. and Weert, Christianus G.},
  year = {1980},
  publisher = {North-Holland Pub. Co.},
  location = {{Amsterdam ; New York : New York}},
  isbn = {978-0-444-85453-7},
  langid = {english},
  pagetotal = {xvii+417},
}

@article{Weaver_1976,
  title = {Reaction Rates in a Relativistic Plasma},
  author = {Weaver, Thomas A.},
  date = {1976-04-01},
  year = 1976,
  journal = {\pra},
  volume = {13},
  number = {4},
  pages = {1563--1569},
  issn = {0556-2791},
  doi = {10.1103/PhysRevA.13.1563},
       adsurl = {https://ui.adsabs.harvard.edu/abs/1976PhRvA..13.1563W},
}

@book{Weekes_2003,
  title = {Very High Energy Gamma-Ray Astronomy},
  author = {Weekes, Trevor C.},
  year = {2003},
  series = {Series in \aap},
  publisher = {{Institute of Physics Pub.}},
  location = {{Bristol ;}},
  isbn = {978-1-4200-3319-9},
}

@article{HoerbeEtAl_2020,
  title = {On the Relative Importance of Hadronic Emission Processes along the Jet Axis of {{Active Galactic Nuclei}}},
  author = {Hoerbe, Mario R. and Morris, Paul J. and Cotter, Garret and Tjus, Julia Becker},
  date = {2020-08-11},
  year = 2020,
  journal = {\mnras},
  volume = {496},
  number = {3},
  eprint = {2006.05140},
  eprinttype = {arxiv},
  eprintclass = {astro-ph},
  pages = {2885--2901},
  issn = {0035-8711, 1365-2966},
  doi = {10.1093/mnras/staa1650},
       adsurl = {https://ui.adsabs.harvard.edu/abs/2020MNRAS.496.2885H},
}

@article{KantzasEtAl_2023,
	title = {Exploring the role of composition and mass loading on the properties of hadronic jets},
	volume = {520},
	issn = {0035-8711, 1365-2966},
	doi = {10.1093/mnras/stad521},
	number = {4},
	journal = {\mnras},
	author = {Kantzas, D and Markoff, S and Lucchini, M and Ceccobello, C and Chatterjee, K},
	month = feb,
	year = {2023},
	pages = {6017--6039},
adsurl = {https://ui.adsabs.harvard.edu/abs/2023MNRAS.520.6017K},
}

@article{LucchiniEtAl_2022,
  title = {Bhjet: A Public Multizone, Steady State Jet + Thermal Corona Spectral Model},
  shorttitle = {Bhjet},
  author = {Lucchini, M and Ceccobello, C and Markoff, S and Kini, Y and Chhotray, A and Connors, R M T and Crumley, P and Falcke, H and Kantzas, D and Maitra, D},
  date = {2022-12-21},
  year = 2022,
  journal = {\mnras},
  volume = {517},
  number = {4},
  pages = {5853--5881},
  issn = {0035-8711},
  doi = {10.1093/mnras/stac2904}
}

@article{Potter_2018,
  title = {Modelling Blazar Flaring Using a Time-Dependent Fluid Jet Emission Model – an Explanation for Orphan Flares and Radio Lags},
  author = {Potter, William J.},
  date = {2018-01-21},
  year = 2018,
  journal = {\mnras},
  volume = {473},
  number = {3},
  pages = {4107--4121},
  issn = {0035-8711},
  doi = {10.1093/mnras/stx2371},
       adsurl = {https://ui.adsabs.harvard.edu/abs/2018MNRAS.473.4107P},
}

@article{McKinneyGammie_2004,
  title = {A {{Measurement}} of the {{Electromagnetic Luminosity}} of a {{Kerr Black Hole}}},
  author = {McKinney, Jonathan C. and Gammie, Charles F.},
  date = {2004-08-20},
  year = {2004},
  journal = {\apj},
  volume = {611},
  number = {2},
  pages = {977},
  issn = {0004-637X},
  doi = {10.1086/422244},
       adsurl = {https://ui.adsabs.harvard.edu/abs/2004ApJ...611..977M},
}

@article{McKinney_2006,
  title = {General Relativistic Magnetohydrodynamic Simulations of the Jet Formation and Large-Scale Propagation from Black Hole Accretion Systems},
  author = {McKinney, Jonathan C.},
  date = {2006-06-01},
  year = {2006},
  journal = {\mnras},
  volume = {368},
  number = {4},
  pages = {1561--1582},
  issn = {0035-8711},
  doi = {10.1111/j.1365-2966.2006.10256.x}
}

@article{ZachariasEtAl_2022,
	title = {{ExHaLe}-jet: an extended hadro-leptonic jet model for blazars – {I}. {Code} description and initial results},
	volume = {512},
	issn = {0035-8711},
	shorttitle = {{ExHaLe}-jet},
	doi = {10.1093/mnras/stac754},
	number = {3},
	urldate = {2023-07-10},
	journal = {\mnras},
	author = {Zacharias, M and Reimer, A and Boisson, C and Zech, A},
	month = may,
	year = {2022},
	pages = {3948--3971},
       adsurl = {https://ui.adsabs.harvard.edu/abs/2022MNRAS.512.3948Z},
}

@article{KlingerEtAl_2024a,
	title = {{AM3}: {An} {Open}-source {Tool} for {Time}-dependent {Lepto}-hadronic {Modeling} of {Astrophysical} {Sources}},
	volume = {275},
	issn = {0067-0049},
	shorttitle = {{AM3}},
	doi = {10.3847/1538-4365/ad725c},
	number = {1},
	journal = {\apjs},
	author = {Klinger, Marc and Rudolph, Annika and Rodrigues, Xavier and Yuan, Chengchao and Fichet de Clairfontaine, Gaëtan and Fedynitch, Anatoli and Winter, Walter and Pohl, Martin and Gao, Shan},
	month = oct,
	year = {2024},
	pages = {4},
       adsurl = {https://ui.adsabs.harvard.edu/abs/2024ApJS..275....4K},
}

@article{CerrutiEtAl_2024,
       author = {{Cerruti}, Matteo and {Rudolph}, Annika and {Petropoulou}, Maria and {B{\"o}ttcher}, Markus and {Stathopoulos}, Stamatios I. and {Oikonomou}, Foteini and {Dimitrakoudis}, Stavros and {Dmytriiev}, Anton and {Gao}, Shan and {Inoue}, Susumu and {Mastichiadis}, Apostolos and {Murase}, Kohta and {Reimer}, Anita and {Robinson}, Joshua and {Rodrigues}, Xavier and {Winter}, Walter and {Zech}, Andreas},
        title = "{A Comprehensive Hadronic Code Comparison for Active Galactic Nuclei}",
      journal = {arXiv:2411.14218},
         year = 2024,
        month = nov,
          doi = {10.48550/arXiv.2411.14218},
archivePrefix = {arXiv},
       eprint = {2411.14218},
 primaryClass = {astro-ph.HE},
}

@article{AksenovEtAl_2010,
	title = {Pair plasma relaxation time scales},
	volume = {81},
	doi = {10.1103/PhysRevE.81.046401},
	number = {4},
	urldate = {2024-01-26},
	journal = {Phys. Rev. E},
	author = {Aksenov, A. G. and Ruffini, R. and Vereshchagin, G. V.},
	month = apr,
	year = {2010},
	pages = {046401},
    adsurl = {https://ui.adsabs.harvard.edu/abs/2010PhRvE..81d6401A}
}

@article{AksenovEtAl_2007,
	title = {{Thermalization} of {Nonequilibrium} {Electron}-{Positron}-{Photon} {Plasmas}},
	volume = {99},
	doi = {10.1103/PhysRevLett.99.125003},
	number = {12},
	urldate = {2024-01-26},
	journal = {Phys. Rev. Lett.},
	author = {Aksenov, A. G. and Ruffini, R. and Vereshchagin, G. V.},
	month = sep,
	year = {2007},
	pages = {125003},
    adsurl = {https://ui.adsabs.harvard.edu/abs/2007PhRvL..99l5003A}
}

@article{AksenovEtAl_2004,
	title = {Structure of {Pair} {Winds} from {Compact} {Objects} with {Application} to {Emission} from {Hot} {Bare} {Strange} {Stars}},
	volume = {609},
	issn = {0004-637X},
	doi = {10.1086/421006},
	number = {1},
	journal = {\apj},
	author = {Aksenov, A. G. and Milgrom, M. and Usov, V. V.},
	month = jul,
	year = {2004},
	pages = {363},
    adsurl = {https://ui.adsabs.harvard.edu/abs/2004ApJ...609..363A}
}

@article{Webb_1985,
	title = {Relativistic {Transport} {Theory} for {Cosmic} {Rays}},
	volume = {296},
	issn = {0004-637X},
	adsurl = {https://ui.adsabs.harvard.edu/abs/1985ApJ...296..319W},
	doi = {10.1086/163451},
	urldate = {2024-07-19},
	journal = {\apj},
	author = {Webb, G. M.},
	month = sep,
	year = {1985},
	pages = {319},
}

@article{Webb_1989a,
	title = {The diffusion approximation and transport theory for cosmic rays in relativistic flows},
	volume = {340},
	issn = {0004-637X, 1538-4357},
	adsurl = {http://adsabs.harvard.edu/doi/10.1086/167462},
	doi = {10.1086/167462},
	urldate = {2024-07-19},
	journal = {\apj},
	author = {Webb, G. M.},
	month = may,
	year = {1989},
	pages = {1112},
}

@article{Lindquist_1966,
	title = {Relativistic transport theory},
	volume = {37},
	issn = {0003-4916},
	doi = {10.1016/0003-4916(66)90207-7},
	number = {3},
	journal = {Ann. Phys. (N. Y.)},
	author = {Lindquist, Richard W.},
	month = may,
	year = {1966},
	pages = {487--518},
       adsurl = {https://ui.adsabs.harvard.edu/abs/1966AnPhy..37..487L},
}

@article{KrumholzEtAl_2022,
	title = {Cosmic ray interstellar propagation tool using {Itô} {Calculus}: software for simultaneous calculation of cosmic ray transport and observational signatures},
	volume = {517},
	issn = {0035-8711, 1365-2966},
	shorttitle = {Cosmic ray interstellar propagation tool using {Itô} {Calculus}},
	doi = {10.1093/mnras/stac2712},
	number = {1},
	journal = {\mnras},
	author = {Krumholz, Mark R. and Crocker, Roland M. and Sampson, Matt L.},
	month = oct,
	year = {2022},
	pages = {1355--1380},
       adsurl = {https://ui.adsabs.harvard.edu/abs/2022MNRAS.517.1355K},
}

@article{KrumholzEtAl_2024,
       author = {Krumholz, Mark R. and Crocker, Roland M. and Bahramian, Arash and Bordas, Pol},
        title = "{Teraelectronvolt gamma-ray emission near globular cluster Terzan 5 as a probe of cosmic ray transport}",
      journal = {Nature Astronomy},
         year = 2024,
        month = oct,
       volume = {8},
       number = {10},
        pages = {1284-1293},
          doi = {10.1038/s41550-024-02337-1},
archivePrefix = {arXiv},
       eprint = {2406.18160},
 primaryClass = {astro-ph.HE},
       adsurl = {https://ui.adsabs.harvard.edu/abs/2024NatAs...8.1284K}
}

@article{EverettCotter_2024,
	title = {Computational forms for binary particle interactions at different levels of anisotropy},
	volume = {3},
	issn = {2752-8200},
	doi = {10.1093/rasti/rzae036},
	number = {1},
	journal = {RASTI},
	author = {Everett, Christopher N and Cotter, Garret},
	month = jan,
	year = {2024},
	pages = {548--555},
       adsurl = {https://ui.adsabs.harvard.edu/abs/2024RASTI...3..548E},
}

@article{ChatterjeeEtAl_2019a,
	title = {Accelerating {AGN} jets to parsec scales using general relativistic {MHD} simulations},
	volume = {490},
	issn = {0035-8711},
	url = {https://doi.org/10.1093/mnras/stz2626},
	doi = {10.1093/mnras/stz2626},
	number = {2},
	journal = {\mnras},
	author = {Chatterjee, K and Liska, M and Tchekhovskoy, A and Markoff, S B},
	month = dec,
	year = {2019},
	pages = {2200--2218},
    adsurl = {https://ui.adsabs.harvard.edu/abs/2019MNRAS.490.2200C},
}

@inproceedings{EhlersJ_1971,
  title = {General {{Relativity}} and {{Kinetic Theory}}},
  booktitle = {General {{Relativity}} and {{Cosmology}}},
  author = {Ehlers, J},
  year = 1971,
  pages = {1--70},
publisher = {Academic Press},
editor = {Sachs, R. K.},
       adsurl = {https://ui.adsabs.harvard.edu/abs/1971grc..conf....1E},
}

@article{ShibataEtAl_2014,
	title = {Conservative form of {Boltzmann}'s equation in general relativity},
	volume = {89},
	doi = {10.1103/PhysRevD.89.084073},
	number = {8},
	journal = {\prd},
	author = {Shibata, Masaru and Nagakura, Hiroki and Sekiguchi, Yuichiro and Yamada, Shoichi},
	month = apr,
	year = {2014},
	pages = {084073},
       adsurl = {https://ui.adsabs.harvard.edu/abs/2014PhRvD..89h4073S},
}

@article{CardallMezzacappa_2003,
	title = {Conservative formulations of general relativistic kinetic theory},
	volume = {68},
	url = {https://link.aps.org/doi/10.1103/PhysRevD.68.023006},
	doi = {10.1103/PhysRevD.68.023006},
	number = {2},
	journal = {\prd},
	author = {Cardall, Christian Y. and Mezzacappa, Anthony},
	month = jul,
	year = {2003},
	pages = {023006},
    adsurl = {https://ui.adsabs.harvard.edu/abs/2003PhRvD..68b3006C},
}

@article{Steeb_1979,
	title = {Generalized liouville equation, entropy, and dynamic systems containing limit cycles},
	volume = {95},
	issn = {0378-4371},
	doi = {10.1016/0378-4371(79)90050-5},
	number = {1},
	journal = {Phys. A: Stat. Mech. Appl.},
	author = {Steeb, W.},
	month = jan,
	year = {1979},
	pages = {181--190},
       adsurl = {https://ui.adsabs.harvard.edu/abs/1979PhyA...95..181S},
}

@article{Steeb_1980,
	title = {A comment on the generalized {Liouville} equation},
	volume = {10},
	issn = {1572-9516},
	doi = {10.1007/BF00708744},
	number = {5},
	journal = {Found. Phys.},
	author = {Steeb, W.},
	month = jun,
	year = {1980},
	pages = {485--493},
       adsurl = {https://ui.adsabs.harvard.edu/abs/1980FoPh...10..485S},
}

@article{Hakim_1967,
	title = {Remarks on {Relativistic} {Statistical} {Mechanics}. {I}},
	volume = {8},
	issn = {0022-2488},
	doi = {10.1063/1.1705347},
	number = {6},
	journal = {J. Math. Phys.},
	author = {Hakim, Rémi},
	month = jun,
	year = {1967},
	pages = {1315--1344},
   adsurl = {https://ui.adsabs.harvard.edu/abs/1967JMP.....8.1315H},
}

@article{Hakim_1967a,
	title = {Remarks on {Relativistic} {Statistical} {Mechanics}. {II}. {Hierarchies} for the {Reduced} {Densities}},
	volume = {8},
	issn = {0022-2488},
	url = {https://doi.org/10.1063/1.1705351},
	doi = {10.1063/1.1705351},
	number = {7},
	journal = {J. Math. Phys.},
	author = {Hakim, Rémi},
	month = jul,
	year = {1967},
	pages = {1379--1400},
       adsurl = {https://ui.adsabs.harvard.edu/abs/1967JMP.....8.1379H},
}

@book{Cercignani_2002,
	address = {Basel},
	series = {Progress in mathematical physics},
  volume = {22},
	title = {The relativistic {Boltzmann} equation: theory and applications},
	isbn = {978-3-7643-6693-3},
	shorttitle = {The relativistic {Boltzmann} equation},
	publisher = {Birkhäuser},
	author = {Cercignani, Carlo},
	collaborator = {Kremer, Gilberto Medeiros},
	year = {2002},
}

@book{Cartan_1922,
  title = {Leçons sur les invariants intégraux},
  author = {Cartan, Elie},
  year = 1922,
  publisher = {Hermann},
  location = {Paris},
  langid = {fre},
  pagetotal = {x+210},
}

@article{Donder_1911,
  title = {Sur Les Invariants Intégraux Relatifs et Leurs Applications à La Physique Mathématique},
  author = {De Donder, Théophile},
  year = {1911},
  journal = {Bulletin de la Classe des sciences. Académie royale de Belgique},
  shortjournal = {Bull. Cl. sci., Acad. r. Belg},
  number = {1},
  pages = {50--70},
  publisher = {Palaies des Académies},
  url = {https://www.biodiversitylibrary.org/page/59013462}
}

@phdthesis{Bichteler_1965,
	title = {Beiträge zur relativistischen kinetischen {Gastheorie}.},
	school = {Inaug.-Diss.--Hamburg.},
	author = {Bichteler, Klaus},
	year = {1965},
}

@book{Lifshits_2008,
	series = {Course of theoretical physics ; v. 10},
	title = {Physical kinetics},
	isbn = {978-0-7506-2635-4},
	publisher = {Elsevier},
	author = {Lifshits, E. M.},
	collaborator = {Pitaevskiĭ, L. P.},
	year = {2008},
}

@article{BlandfordEichler_1987,
	title = {Particle acceleration at astrophysical shocks: {A} theory of cosmic ray origin},
	volume = {154},
	issn = {0370-1573},
	shorttitle = {Particle acceleration at astrophysical shocks},
	url = {https://www.sciencedirect.com/science/article/pii/0370157387901347},
	doi = {10.1016/0370-1573(87)90134-7},
	number = {1},
	urldate = {2025-05-16},
	journal = {\physrep},
	author = {Blandford, Roger and Eichler, David},
	month = oct,
	year = {1987},
	pages = {1--75},
    adsurl = {https://ui.adsabs.harvard.edu/abs/1987PhR...154....1B}
}

@misc{Helein_2023,
	title = {Answer to ``{Is} {Cartan}'s magic formula'' due to Élie or {Henri}?"},
	shorttitle = {Answer to "{Is}``{Cartan}'s magic formula'' due to Élie or {Henri}?},
	url = {https://mathoverflow.net/a/438297},
	urldate = {2025-01-07},
	journal = {MathOverflow},
	author = {Hélein, Frédéric},
	month = jan,
	year = {2023},
}

@book{Abraham_1905,
	address = {Leipzig},
	title = {Theorie der elektrizität. {Zweiter} {Band}: {Elektromagnetische} {Theorie} der {Strahlung}},
	publisher = {B.G. Teubner},
	author = {Abraham, Max},
	collaborator = {Föppl, Aug (August)},
	year = {1905},
}

@book{Lorentz_1892,
	address = {Netherlands},
	title = {La théorie électromagnétique de {Maxwell} et son application aux corps mouvants par {H}.{A}. {Lorentz}},
	urldate = {2025-08-07},
	publisher = {E.J. Brill},
	author = {Lorentz, H. A.},
	year = {1892},
}

@article{Dirac_1938,
	title = {Classical theory of radiating electrons},
	volume = {167},
	doi = {10.1098/rspa.1938.0124},
	number = {929},
	urldate = {2024-04-05},
	journal = {Proc. R. Soc. A},
	author = {Dirac, Paul Adrien Maurice},
	year = {1938},
	pages = {148--169},
       adsurl = {https://ui.adsabs.harvard.edu/abs/1938RSPSA.167..148D},
}

@article{PoissonEtAl_2011,
	title = {The {Motion} of {Point} {Particles} in {Curved} {Spacetime}},
	volume = {14},
	issn = {1433-8351},
	doi = {10.12942/lrr-2011-7},
	number = {1},
	urldate = {2024-04-05},
	journal = {Living Rev. Relativ.},
	author = {Poisson, Eric and Pound, Adam and Vega, Ian},
	month = sep,
	year = {2011},
	pages = {7},
       adsurl = {https://ui.adsabs.harvard.edu/abs/2011LRR....14....7P},
}

@article{GrallaEtAl_2009,
	title = {Rigorous derivation of electromagnetic self-force},
	volume = {80},
	doi = {10.1103/PhysRevD.80.024031},
	number = {2},
	urldate = {2024-04-05},
	journal = {\prd},
	author = {Gralla, Samuel E. and Harte, Abraham I. and Wald, Robert M.},
	month = jul,
	year = {2009},
	pages = {024031},
       adsurl = {https://ui.adsabs.harvard.edu/abs/2009PhRvD..80b4031G},
}

@article{GellesEtAl_2025,
	title = {Signatures of {Black} {Hole} {Spin} and {Plasma} {Acceleration} in {Jet} {Polarimetry}},
	volume = {981},
	issn = {0004-637X},
	doi = {10.3847/1538-4357/adb1aa},
	number = {2},
	journal = {\apj},
	author = {Gelles, Z. and Chael, A. and Quataert, E.},
	year = {2025},
	pages = {204},
       adsurl = {https://ui.adsabs.harvard.edu/abs/2025ApJ...981..204G},
}

@article{TsunetoeEtAl_2025,
	title = {Limb-brightened {Jet} in {M87} from {Anisotropic} {Nonthermal} {Electrons}},
	volume = {984},
	issn = {0004-637X},
	doi = {10.3847/1538-4357/adc37a},
	language = {en},
	number = {1},
	journal = {\apj},
	author = {Tsunetoe, Yuh and Pesce, Dominic W. and Narayan, Ramesh and Chael, Andrew and Gelles, Zachary and Gammie, Charles and Quataert, Eliot and Palumbo, Daniel},
	month = apr,
	year = {2025},
	pages = {35},
       adsurl = {https://ui.adsabs.harvard.edu/abs/2025ApJ...984...35T},
}

@article{ChaelEtAl_2023,
	title = {Black {Hole} {Polarimetry} {I}: {A} {Signature} of {Electromagnetic} {Energy} {Extraction}},
	volume = {958},
	issn = {0004-637X, 1538-4357},
	shorttitle = {Black {Hole} {Polarimetry} {I}},
	doi = {10.3847/1538-4357/acf92d},
	number = {1},
	journal = {\apj},
	author = {Chael, Andrew and Lupsasca, Alexandru and Wong, George N. and Quataert, Eliot},
	year = {2023},
	pages = {65},
    adsurl = {https://ui.adsabs.harvard.edu/abs/2023ApJ...958...65C},
}

@article{SarkarEtAl_2019,
	title = {Dissecting the {Compton} scattering kernel {I}: {Isotropic} media},
	volume = {490},
	issn = {0035-8711, 1365-2966},
	doi = {10.1093/mnras/stz2794},
	number = {3},
	journal = {\mnras},
	author = {Sarkar, Abir and Chluba, Jens and Lee, Elizabeth},
	year = {2019},
	pages = {3705--3726},
       adsurl = {https://ui.adsabs.harvard.edu/abs/2019MNRAS.490.3705S},
}

@article{BlumenthalGould_1970,
	title = {Bremsstrahlung, {Synchrotron} {Radiation}, and {Compton} {Scattering} of {High}-{Energy} {Electrons} {Traversing} {Dilute} {Gases}},
	volume = {42},
	issn = {0034-6861},
	doi = {10.1103/RevModPhys.42.237},
	number = {2},
	journal = {Rev. Mod. Phys.},
	author = {Blumenthal, George R. and Gould, Robert J.},
	month = apr,
	year = {1970},
	pages = {237--270},
    adsurl = {https://ui.adsabs.harvard.edu/abs/1970RvMP...42..237B}
}

@article{BlandfordEtAl_2019,
	title = {Relativistic {Jets} from {Active} {Galactic} {Nuclei}},
	volume = {57},
	doi = {10.1146/annurev-astro-081817-051948},
	number = {1},
	journal = {\araa},
	author = {Blandford, Roger and Meier, David and Readhead, Anthony},
	year = {2019},
	pages = {467--509},
    adsurl = {https://ui.adsabs.harvard.edu/abs/2019ARA&A..57..467B}
}

@article{PadovaniEtAl_2017,
	title = {Active {Galactic} {Nuclei}: what's in a name?},
	volume = {25},
	issn = {0935-4956, 1432-0754},
	shorttitle = {Active {Galactic} {Nuclei}},
	doi = {10.1007/s00159-017-0102-9},
	number = {1},
	journal = {\aapr},
	author = {Padovani, P. and Alexander, D. M. and Assef, R. J. and Marco, B. De and Giommi, P. and Hickox, R. C. and Richards, G. T. and Smolcic, V. and Hatziminaoglou, E. and Mainieri, V. and Salvato, M.},
	month = nov,
	year = {2017},
       adsurl = {https://ui.adsabs.harvard.edu/abs/2017A&ARv..25....2P},
}

@article{BottcherEtAl_2013,
	title = {Leptonic and {Hadronic} {Modeling} of {Fermi}-detected {Blazars}},
	volume = {768},
	issn = {0004-637X},
	doi = {10.1088/0004-637X/768/1/54},
	journal = {\apj},
	author = {Böttcher, M. and Reimer, A. and Sweeney, K. and Prakash, A.},
	month = may,
	year = {2013},
	pages = {54},
    adsurl = {https://ui.adsabs.harvard.edu/abs/2013ApJ...768...54B},
}

@article{klingerEtAl_2024,
	title = {Leptohadronic {Scenarios} for {TeV} {Extensions} of {Gamma}-{Ray} {Burst} {Afterglow} {Spectra}},
	volume = {977},
	issn = {0004-637X},
	doi = {10.3847/1538-4357/ad9392},
	language = {en},
	number = {2},
	journal = {\apj},
	author = {Klinger, Marc and Yuan, Chengchao and Taylor, Andrew M. and Winter, Walter},
	month = dec,
	year = {2024},
	pages = {242},
       adsurl = {https://ui.adsabs.harvard.edu/abs/2024ApJ...977..242K},
}

@article{MarkoffEtAl_2001,
	title = {A jet model for the broadband spectrum of {XTE} {J1118}+480 - {Synchrotron} emission from radio to {X}-rays in the {Low}/{Hard} spectral state},
	volume = {372},
	issn = {0004-6361, 1432-0746},
	doi = {10.1051/0004-6361:20010420},
	number = {2},
	urldate = {2023-09-25},
	journal = {\aap},
	author = {Markoff, S. and Falcke, H. and Fender, R.},
	month = jun,
	year = {2001},
	pages = {L25--L28},
       adsurl = {https://ui.adsabs.harvard.edu/abs/2001A&A...372L..25M},
}

@article{Marscher_2014,
	title = {Turbulent, {Extreme} {Multi}-zone {Model} for {Simulating} {Flux} and {Polarization} {Variability} in {Blazars}},
	volume = {780},
	issn = {0004-637X},
	doi = {10.1088/0004-637X/780/1/87},
	journal = {\apj},
	author = {Marscher, Alan P.},
	month = jan,
	year = {2014},
	pages = {87},
       adsurl = {https://ui.adsabs.harvard.edu/abs/2014ApJ...780...87M},
}

@article{KantzasEtAl_2021a,
	title = {A new lepto-hadronic model applied to the first simultaneous multiwavelength data set for {Cygnus} {X}–1},
	volume = {500},
	issn = {0035-8711},
	url = {https://doi.org/10.1093/mnras/staa3349},
	doi = {10.1093/mnras/staa3349},
	number = {2},
	journal = {\mnras},
	author = {Kantzas, D and Markoff, S and Beuchert, T and Lucchini, M and Chhotray, A and Ceccobello, C and Tetarenko, A J and Miller-Jones, J C A and Bremer, M and Garcia, J A and Grinberg, V and Uttley, P and Wilms, J},
	month = jan,
	year = {2021},
	pages = {2112--2126},
       adsurl = {https://ui.adsabs.harvard.edu/abs/2021MNRAS.500.2112K},
}

@article{Liouville_1838,
	title = {Note sur la {Théorie} de la {Variation} des constantes arbitraires.},
	issn = {0021-7874},
	journal = {\href{https://eudml.org/doc/234417}{J. Math. Pures Appl.}},
	author = {Liouville, J.},
	year = {1838},
	pages = {342--349},
}

@article{EllisEtAl_1983,
	title = {Anisotropic solutions of the {Einstein}-{Boltzmann} equations: {I}. {General} formalism},
	volume = {150},
	issn = {0003-4916},
	doi = {10.1016/0003-4916(83)90023-4},
	number = {2},
	journal = {Ann. Phys. (N. Y.)},
	author = {Ellis, G. F. R and Matravers, D. R and Treciokas, R},
	month = oct,
	year = {1983},
	pages = {455--486},
       adsurl = {https://ui.adsabs.harvard.edu/abs/1983AnPhy.150..455E},
}

@article{EllisEtAl_1983a,
	title = {Anisotropic solutions of the {Einstein}-{Boltzmann} equations. {II}. {Some} exact properties of the equations},
	volume = {150},
	issn = {0003-4916},
	doi = {10.1016/0003-4916(83)90024-6},
	number = {2},
	journal = {Ann. Phys. (N. Y.)},
	author = {Ellis, G. F. R and Treciokas, R and Matravers, D. R},
	month = oct,
	year = {1983},
	pages = {487--503},
       adsurl = {https://ui.adsabs.harvard.edu/abs/1983AnPhy.150..487E},
}

@article{WhiteheadMatthews_2023,
	title = {Studying the link between radio galaxies and {AGN} fuelling with relativistic hydrodynamic simulations of flickering jets},
	volume = {523},
	issn = {0035-8711},
	doi = {10.1093/mnras/stad1582},
	number = {2},
	journal = {\mnras},
	author = {Whitehead, Henry W and Matthews, James H},
	month = aug,
	year = {2023},
	pages = {2478--2497},
       adsurl = {https://ui.adsabs.harvard.edu/abs/2023MNRAS.523.2478W},
}

@article{GambaRjasanow_2018,
	title = {Galerkin–{Petrov} approach for the {Boltzmann} equation},
	volume = {366},
	issn = {0021-9991},
	doi = {10.1016/j.jcp.2018.04.017},
	urldate = {2025-08-12},
	journal = {J. Comput. Phys.},
	author = {Gamba, Irene M. and Rjasanow, Sergej},
	month = aug,
	year = {2018},
	pages = {341--365},
       adsurl = {https://ui.adsabs.harvard.edu/abs/2018JCoPh.366..341G},
}

@article{CockburnShu_2001,
	title = {Runge–{Kutta} {Discontinuous} {Galerkin} {Methods} for {Convection}-{Dominated} {Problems}},
	volume = {16},
	issn = {1573-7691},
	doi = {10.1023/A:1012873910884},
	number = {3},
	urldate = {2025-08-12},
	journal = {J. Sci. Comput.},
	author = {Cockburn, Bernardo and Shu, Chi-Wang},
	month = sep,
	year = {2001},
	pages = {173--261},
}

@article{GasparyanEtAl_2022,
	title = {Time-dependent lepto-hadronic modelling of the emission from blazar jets with {SOPRANO}: the case of {TXS} 0506 + 056, {3HSP} {J095507}.9 + 355101, and {3C} 279},
	volume = {509},
	issn = {0035-8711},
	shorttitle = {Time-dependent lepto-hadronic modelling of the emission from blazar jets with {SOPRANO}},
	doi = {10.1093/mnras/stab2688},
	urldate = {2025-08-12},
	journal = {\mnras},
	author = {Gasparyan, S. and Bégué, D. and Sahakyan, N.},
	month = jan,
	year = {2022},
	pages = {2102--2121},
       adsurl = {https://ui.adsabs.harvard.edu/abs/2022MNRAS.509.2102G},
}

@article{Boltzmann_1872,
	title = {Weitere {Studien} über das {Wärmegleichgewicht} unter {Gas} molekülen.},
	volume = {66},
	url = {https://babel.hathitrust.org/cgi/pt?id=hvd.32044093294999&view=1up&seq=832},
	journal = {Sitzungsberichte der Kaiserlichen Akademie der Wissenschaften},
	author = {Boltzmann, Ludwig},
	year = {1872},
	pages = {275--390},
}

@article{PaperII,
	title={DIPLODOCUS II: Implementation of transport equations and test cases relevant to micro-scale physics of jetted astrophysical sources}, 
	author={Christopher N. Everett and Marc Klinger-Plaisier and Garret Cotter},
	year={2025},
    journal = {arXiv:2510.12505},
	eprint={2510.12505},
	archivePrefix={arXiv},
	primaryClass={astro-ph.HE},
    shorthand = {Paper II},
    doi = {10.48550/arXiv.2510.12505},
    note = {(Paper II)}
}

@unpublished{PaperIII,
    year = {in prep.},
    author = {Everett, Christopher N. and Cotter, Garret},
    shorthand = {Paper III},
    note = {(Paper III)}
}

@article{Stigler_1980,
	title = {Stigler's {Law} of {Eponymy}},
	volume = {39},
	copyright = {1980 The New York Academy of Sciences},
	issn = {2164-0947},
	doi = {10.1111/j.2164-0947.1980.tb02775.x},
	number = {1 Series II},
	journal = {Trans. N. Y. Acad. Sci.},
	author = {Stigler, Stephen M.},
	year = {1980},
	pages = {147--157},
}

@article{CeruttiEtAl_2013,
	title = {Simulations of {Particle} {Acceleration} beyond the {Classical} {Synchrotron} {Burnoff} {Limit} in {Magnetic} {Reconnection}: {An} {Explanation} of the {Crab} {Flares}},
	volume = {770},
	issn = {0004-637X},
	shorttitle = {Simulations of {Particle} {Acceleration} beyond the {Classical} {Synchrotron} {Burnoff} {Limit} in {Magnetic} {Reconnection}},
	doi = {10.1088/0004-637X/770/2/147},
	journal = {\apj},
	author = {Cerutti, B. and Werner, G. R. and Uzdensky, D. A. and Begelman, M. C.},
	month = jun,
	year = {2013},
	pages = {147},
    adsurl = {https://ui.adsabs.harvard.edu/abs/2013ApJ...770..147C},
}

@article{MehlhaffEtAl_2024,
	title = {Kinetic simulations and gamma-ray signatures of {Klein}-{Nishina} relativistic magnetic reconnection},
	volume = {527},
	issn = {0035-8711},
	doi = {10.1093/mnras/stad3863},
	journal = {\mnras},
	author = {Mehlhaff, J. and Werner, G. and Cerutti, B. and Uzdensky, D. and Begelman, M.},
	month = feb,
	year = {2024},
	pages = {11587--11626},
       adsurl = {https://ui.adsabs.harvard.edu/abs/2024MNRAS.52711587M},
}

@article{SironiEtAl_2015,
	title = {Relativistic {Shocks}: {Particle} {Acceleration} and {Magnetization}},
	volume = {191},
	issn = {1572-9672},
	shorttitle = {Relativistic {Shocks}},
	doi = {10.1007/s11214-015-0181-8},
	number = {1},
	journal = {\ssr},
	author = {Sironi, L. and Keshet, U. and Lemoine, M.},
	month = oct,
	year = {2015},
	pages = {519--544},
       adsurl = {https://ui.adsabs.harvard.edu/abs/2015SSRv..191..519S},
}

@article{ComissoEtAl_2024,
	title = {Ultra-{High}-{Energy} {Cosmic} {Rays} {Accelerated} by {Magnetically} {Dominated} {Turbulence}},
	volume = {977},
	issn = {2041-8205},
	doi = {10.3847/2041-8213/ad955f},
	number = {1},
	journal = {\apjl},
	author = {Comisso, Luca and Farrar, Glennys R. and Muzio, Marco S.},
	month = dec,
	year = {2024},
	pages = {L18},
    adsurl = {https://ui.adsabs.harvard.edu/abs/2024ApJ...977L..18C},
}

@article{StathopoulosEtAl_2024,
	title = {{LeHaMoC}: {A} versatile time-dependent lepto-hadronic modeling code for high-energy astrophysical sources},
	volume = {683},
	issn = {0004-6361, 1432-0746},
	shorttitle = {{LeHaMoC}},
	doi = {10.1051/0004-6361/202347277},
	journal = {\aap},
	author = {Stathopoulos, S. I. and Petropoulou, M. and Vasilopoulos, G. and Mastichiadis, A.},
	month = mar,
	year = {2024},
	pages = {A225},
       adsurl = {https://ui.adsabs.harvard.edu/abs/2024A&A...683A.225S},
}

@article{GaoEtAl_2017,
	title = {On the {Direct} {Correlation} between {Gamma}-{Rays} and {PeV} {Neutrinos} from {Blazars}},
	volume = {843},
	issn = {0004-637X},
	doi = {10.3847/1538-4357/aa7754},
	journal = {\apj},
	author = {Gao, Shan and Pohl, Martin and Winter, Walter},
	year = {2017},
	pages = {109},
       adsurl = {https://ui.adsabs.harvard.edu/abs/2017ApJ...843..109G},
}

@article{MastichiadisKirk_1995,
	title = {Self-consistent particle acceleration in active galactic nuclei.},
	volume = {295},
	issn = {0004-6361},
	url = {https://ui.adsabs.harvard.edu/abs/1995A&A...295..613M},
	journal = {\aap},
	author = {Mastichiadis, A. and Kirk, J. G.},
	month = mar,
	year = {1995},
	pages = {613},
       adsurl = {https://ui.adsabs.harvard.edu/abs/1995A&A...295..613M},
}

@article{MastichiadisEtAl_2005,
	title = {Spectral and temporal signatures of ultrarelativistic protons in compact sources. {I}. {Effects} of {Bethe}-{Heitler} pair production},
	volume = {433},
	issn = {0004-6361},
	doi = {10.1051/0004-6361:20042161},
	journal = {\aap},
	author = {Mastichiadis, A. and Protheroe, R. J. and Kirk, J. G.},
	month = apr,
	year = {2005},
	pages = {765--776},
      adsurl = {https://ui.adsabs.harvard.edu/abs/2005A&A...433..765M},
}

@article{DimitrakoudisEtAl_2012a,
	title = {The time-dependent one-zone hadronic model. {First} principles},
	volume = {546},
	issn = {0004-6361},
	doi = {10.1051/0004-6361/201219770},
	journal = {\aap},
	author = {Dimitrakoudis, S. and Mastichiadis, A. and Protheroe, R. J. and Reimer, A.},
	month = oct,
	year = {2012},
	pages = {A120},
       adsurl = {https://ui.adsabs.harvard.edu/abs/2012A&A...546A.120D},
}

@article{JimenezFernandezvanEerten_2021,
      title={Katu: a fast open-source full lepto-hadronic kinetic code suitable for Bayesian inference modelling of blazars}, 
      journal = {arXiv:2104.08207},
      author={Bruno Jiménez Fernández and Hendrik van Eerten},
      year={2021},
      eprint={2104.08207},
      archivePrefix={arXiv},
      primaryClass={astro-ph.HE}, 
doi = {10.48550/arXiv.2104.08207}
}

@article{CerrutiEtAl_2015,
	title = {A hadronic origin for ultra-high-frequency-peaked {BL} {Lac} objects},
	volume = {448},
	issn = {0035-8711},
	doi = {10.1093/mnras/stu2691},
	journal = {\mnras},
	author = {Cerruti, M. and Zech, A. and Boisson, C. and Inoue, S.},
	month = mar,
	year = {2015},
	pages = {910--927},
    adsurl = {https://ui.adsabs.harvard.edu/abs/2015MNRAS.448..910C},
}

@article{HahnEtAl_2022,
	title = {{GAMERA}: {Source} modeling in gamma astronomy},
	shorttitle = {{GAMERA}},
	journal = {\href{https://ascl.net/2203.007}{Astrophysics Source Code Library, record ascl:2203.007}},
	author = {Hahn, Joachim and Romoli, Carlo and Breuhaus, Mischa},
	month = mar,
	year = {2022},
	eid = {ascl:2203.007},
    archivePrefix = {ascl},
    eprint = {2203.007},
    adsurl = {https://ui.adsabs.harvard.edu/abs/2022ascl.soft03007H},
}

@article{AharonianAtoyan_1981,
  title = {Compton {{Scattering}} of {{Relativistic Electrons}} in {{Compact X-Ray Sources}}},
  author = {Aharonian, F. A. and Atoyan, A. M.},
  year = 1981,
  month = oct,
  journal = {\apss},
  volume = {79},
  pages = {321--336},
  issn = {0004-640X},
  doi = {10.1007/BF00649428},
  adsurl = {https://ui.adsabs.harvard.edu/abs/1981Ap&SS..79..321A}
}

@article{Brinkmann_1984,
	title = {Compton scattering from isotropic electrons},
	volume = {31},
	issn = {00224073},
	doi = {10.1016/0022-4073(84)90068-2},
	number = {5},
	journal = {\jqsrt},
	author = {Brinkmann, W.},
	month = may,
	year = {1984},
	pages = {417--421},
    adsurl = {https://ui.adsabs.harvard.edu/abs/1984JQSRT..31..417B},
}

@article{MoskalenkoStrong_2000,
	title = {Anisotropic {Inverse} {Compton} {Scattering} in the {Galaxy}},
	volume = {528},
	issn = {0004-637X},
	doi = {10.1086/308138},
	number = {1},
	journal = {\apj},
	author = {Moskalenko, Igor V. and Strong, Andrew W.},
	month = jan,
	year = {2000},
	pages = {357},
       adsurl = {https://ui.adsabs.harvard.edu/abs/2000ApJ...528..357M},
}

@article{KelnerEtAl_2014,
	title = {{THE} {BEAMING} {PATTERN} {OF} {EXTERNAL} {COMPTON} {EMISSION} {FROM} {RELATIVISTIC} {OUTFLOWS}: {THE} {CASE} {OF} {ANISOTROPIC} {DISTRIBUTION} {OF} {ELECTRONS}},
	volume = {785},
	issn = {0004-637X},
	doi = {10.1088/0004-637X/785/2/141},
	number = {2},
	journal = {\apj},
	author = {Kelner, S. R. and Lefa, E. and Rieger, F. M. and Aharonian, F. A.},
	month = apr,
	year = {2014},
	pages = {141},
       adsurl = {https://ui.adsabs.harvard.edu/abs/2014ApJ...785..141K},
}

@article{LaiNg_2023,
	title = {Anisotropic {Photon} and {Electron} {Scattering} without {Ultrarelativistic} {Approximation}},
	volume = {107},
	issn = {2470-0010, 2470-0029},
	doi = {10.1103/PhysRevD.107.063026},
	number = {6},
	journal = {\prd},
	author = {Lai, Anderson C. M. and Ng, Kenny C. Y.},
	month = mar,
	year = {2023},
	pages = {063026},
       adsurl = {https://ui.adsabs.harvard.edu/abs/2023PhRvD.107f3026L},
}

@article{Boula_2022,
  author = {{Boula}, S. and {Mastichiadis}, A.},
  title = "{Expanding one-zone model for blazar emission}",
  journal = {\aap},
  year = 2022,
  month = jan,
  eid = {A20},
  pages = {A20},
  doi = {10.1051/0004-6361/202142126},
  adsurl = {https://ui.adsabs.harvard.edu/abs/2022A&A...657A..20B},
}

@book{Stewart_1971,
  title = {Non-Equilibrium Relativistic Kinetic Theory},
  author = {Stewart, John},
  year = 1971,
  series = {Lecture Notes in Physics},
  number = {10},
  publisher = {Springer-Verlag},
  address = {Berlin},
  isbn = {978-3-540-05652-2},
  langid = {english},
}

@ARTICLE{Talvikki_2019,
       author = {{Hovatta}, Talvikki and {Lindfors}, Elina},
        title = "{Relativistic Jets of Blazars}",
      journal = {\nar},
         year = 2019,
        month = dec,
       volume = {87},
          eid = {101541},
        pages = {101541},
          doi = {10.1016/j.newar.2020.101541},
archivePrefix = {arXiv},
       eprint = {2003.06322},
 primaryClass = {astro-ph.HE},
       adsurl = {https://ui.adsabs.harvard.edu/abs/2019NewAR..8701541H}
}
